\documentclass[11pt]{article}

\usepackage[margin=1in]{geometry}

\usepackage{algorithm}
\usepackage{algpseudocode}
\usepackage{hyperref}

\usepackage{amsmath}
\usepackage{mathtools}
\mathtoolsset{showonlyrefs}
\usepackage{booktabs}   
\usepackage{url}
\usepackage{amssymb}
\usepackage{amsthm}
\newcommand{\field}[1]{\ensuremath{\mathbb{#1}}}
\newcommand{\R}{\ensuremath{\field{R}}} 
\newcommand{\1}{\ensuremath{\mathbf{1}}} 
\newcommand{\I}[1]{\ensuremath{\mathbb{I}_{\left\{#1\right\}}}} 
\newcommand{\PR}{\ensuremath{\mathsf{P}}} 
\newcommand{\E}{\ensuremath{\mathsf{E}}} 

\newcommand{\Fscr}{\ensuremath{\mathcal F}}

\newcommand{\Iscr}{\ensuremath{\mathcal I}}
\newcommand{\Jscr}{\ensuremath{\mathcal J}}

\DeclareMathOperator{\Var}{Var}
\DeclareMathOperator{\Cov}{Cov}

\DeclareMathOperator*{\argmax}{\mathrm{argmax}}

\newtheorem{theorem}{Theorem}

\newtheorem{lemma}{Lemma}

\newtheorem{assumption}{Assumption}

\usepackage{bm}
\usepackage{comment}
\usepackage{caption,subcaption}
\usepackage{float}
\usepackage{enumitem}
\usepackage[dvipsnames]{xcolor}

\usepackage{tikz}
\usetikzlibrary{positioning, arrows.meta}
\usepackage{caption}
\usepackage{array}
\usepackage{pgfplots}
\usepackage{pgfplotstable}
\usepackage{subcaption}
\usepackage{tabularx, longtable}
\pgfplotsset{compat=newest}

\usepackage{natbib}
\bibpunct[, ]{(}{)}{,}{a}{}{,}%
\title{Is Thompson Sampling Susceptible to Algorithmic Collusion?}
\author{Yi Xiong\thanks{School of Information Management and Engineering, Shanghai University of Finance and Economics; xiongyi@mail.shufe.edu.cn} \quad 
	Ningyuan Chen\thanks{The Rotman School of Management, University of Toronto; ningyuan.chen@utoronto.ca} \quad 
	Xuefeng Gao \thanks{Department of Systems Engineering and Engineering Management, The Chinese University of Hong Kong; xfgao@se.cuhk.edu.hk} \quad}

\begin{document}
	\maketitle
	
	\begin{abstract}
			When two players are engaged in a repeated game with unknown payoff matrices, they may use single-agent multi-armed bandit algorithms to choose the actions independent of each other. We show that when the players use Thompson sampling, the game dynamics converges to the Nash equilibrium under a mild assumption on the payoff matrices. Therefore, algorithmic collusion doesn't arise in this case despite the fact that the players do not intentionally deploy competitive strategies. To prove the convergence result, we find that the framework developed in stochastic approximation doesn't apply, because of the sporadic and infrequent updates of the inferior actions and the lack of Lipschitz continuity. We develop a novel sample-path-wise approach to show the convergence. However, when the payoff matrices do not satisfy the assumption, the game may converge to collusive outcomes.
	\end{abstract}
	
\section{Introduction}\label{sec:intro}
The proliferation of learning algorithms in competitive markets has given rise to a significant regulatory and academic concern: \emph{algorithmic collusion}. This phenomenon occurs when multiple firms, each independently using an algorithm to optimize its decisions (e.g., pricing), collectively steer the market towards a collusive outcome with supracompetitive prices, rather than a traditional competitive Nash equilibrium. For instance, in a classic Bertrand competition, firms with full knowledge of the market might settle at a low-price Nash Equilibrium. However, if these firms instead deploy learning algorithms to explore an unknown demand function, they may learn to avoid aggressive price competition, stabilizing at higher prices that benefit all firms at the expense of consumers.

A growing body of research has confirmed that this concern is not merely hypothetical. Several standard single-agent learning algorithms, when deployed by competing players, have been shown to facilitate collusive outcomes. Prominent examples include \emph{Q-learning} \citep{calvano2020artificial,banchio2022artificial,abada2023artificial,meylahn2024does}, \emph{UCB} algorithms \citep{hansen2021frontiers}, and $\epsilon$-greedy algorithms \citep{calvano2021algorithmic,klein2021autonomous,banchio2022artificial,abada2024collusion}. 
This behavior is perhaps unsurprising. These algorithms are designed to learn a stationary environment without other self-interested agents; they are not inherently equipped to handle the non-stationarity introduced by a strategic opponent who is simultaneously learning and adapting. In contrast, algorithms explicitly designed for multi-agent or adversarial settings, which assume the environment can react strategically, reliably guide players toward a Nash equilibrium \citep{cesa2006prediction,mertikopoulos2019learning,meylahn2022learning,cai2023uncoupled,ba2025doubly}. 
Moreover, several papers \citep{bubeck2012best,seldin2014one,zimmert2019optimal} aim to achieve sublinear regret in both stochastic and adversarial environments. Such algorithms may perceive the price fluctuations of a competing algorithm as an unpredictable and non-stationary ``adversary" and are more likely to switch to a strategy that involves more exploration of different actions to break the collusion.

Despite the theoretical appeal of these robust, multi-agent algorithms, the behavior of simpler, single-agent algorithms remains a central concern for two critical reasons. First, their simplicity, scalability, and efficiency make them attractive for real-world deployment, especially when a firm perceives its environment as being more stochastic than strategic. A firm might plausibly deploy a standard reinforcement learning tool without fully accounting for its interactive effects with competitors' algorithms. Second, from a firm's perspective, converging to a supracompetitive outcome is often more profitable than converging to a Nash equilibrium. The incentive to use an algorithm that might lead to collusion, even if unintentionally, is therefore strong.

Conspicuously absent from this line of literature is one of the most celebrated and widely deployed learning algorithms: \emph{Thompson sampling}, 
a popular Bayesian algorithm in multi-armed bandits and reinforcement learning \citep{russo2018tutorial}.
It is also known as posterior sampling.
Recognized as a state-of-the-art method, Thompson sampling's wide adoption in practice is driven by both its superior empirical performance and its straightforward implementation, making it highly effective for tasks such as advertisement selection, clinical trials, and personalized recommendation \citep{chapelle2011empirical,scott2015multi}.
Known for its Bayesian motivation, impressive empirical performance, and flexibility in handling a wide range of problem setups, its behavior in a competitive game setting remains largely an open question. 
This represents a critical gap in our understanding of algorithmic collusion. This study, therefore, strives to answer the following question:
\begin{quote}
	\centering
	\emph{Is Thompson sampling susceptible to algorithmic collusion?}
\end{quote}

\subsection{A Motivating Example}\label{sec:motivating-exp}
To frame this question more concretely, we consider a simple numerical experiment based on the classic Prisoner's Dilemma. Two players each have two actions, which we can label ``defect" (action 1) and ``cooperate" (action 2). The payoff matrices are:
\begin{align}
	A = \begin{pmatrix}
		0.2 & 5 \\
		0.1 & 4
	\end{pmatrix}, \quad
	B = \begin{pmatrix}
		0.2 & 0.1 \\
		5 & 4
	\end{pmatrix}.
\end{align} 
That is, if Player 1 chooses action $i$ and Player 2 chooses action $j$, the means of their respective random payoffs are $A_{ij}$ and $B_{ij}$.
The unique pure-strategy Nash equilibrium is $(1,1)$ (defect, defect), with a payoff of $(0.2, 0.2)$. However, the outcome $(2,2)$ (cooperate, cooperate) is mutually preferable, yielding a higher payoff of $(4,4)$. This latter outcome represents the collusive point.

We simulate a scenario where the players are unaware of these payoff matrices and must learn how to act based only on the rewards they observe. Each player independently employs a Thompson sampling algorithm to maximize their own cumulative reward. 
Crucially, each player doesn't consider the strategic nature and models the problem as a simple two-armed bandit task. 
The two available actions (``defect'' and ``cooperate'') are treated as two independent arms of a bandit. This creates a non-stationary reward environment for each player, as their opponent is simultaneously learning and adapting their strategy.
In each round $n$, Player 1 chooses action 1 with some probability $\varphi_{1,n}$ and Player 2 chooses action 1 with probability $\psi_{1,n}$, based on their respective histories. 
Convergence to the Nash equilibrium would mean $(\varphi_{1,n}, \psi_{1,n}) \to (1,1)$, while convergence to the collusive outcome would mean $(\varphi_{1,n}, \psi_{1,n}) \to (0,0)$.
More details about this example can be found in Section~\ref{sec:experiments}.
We simulate 500 independent paths and take the average of $\varphi$ and $\psi$ among the paths in each round.
The trajectory of the average probability of both players taking action 1 is presented in Figure~\ref{fig:ts-collude}.
\begin{figure}
	\centering
	\begin{tikzpicture}
		\begin{axis}[
			xlabel={Time},
			ylabel={Average Probability},
			xlabel style={font=\small},
			ylabel style={font=\small},
			xmin=1, xmax=1000000,
			ymin=0, ymax=1,
			xmode=log,
			log basis x=10,
			xtick={1,100,10000,1000000},
			xticklabels={1,$10^2$,$10^4$,$10^6$},
			ytick={0,0.2,0.4,0.6,0.8,1},
			xticklabel style={font=\small},
			yticklabel style={font=\small},
			legend pos=south east,
			legend style={font=\small},
			grid=minor
			]
			\addplot+[solid, thick,mark=none] table [x=Time, y=Phi, col sep=comma] {Data/TS2-500.csv};
			\addlegendentry{Player 1 Action 1}
			\addplot+[dashed, thick,mark=none] table [x=Time, y=Psi, col sep=comma] {Data/TS2-500.csv};
			\addlegendentry{Player 2 Action 1}
		\end{axis}
	\end{tikzpicture}
	\caption{The average probability that two players choose action 1 (defect) with payoff matrices $(A,B)$ over 500 sample paths.}
	\label{fig:ts-collude}
\end{figure}

As one can observe, the game doesn't always converge to the Nash equilibrium.
In fact, for around 22.6\% of the runs, the game converges to the collusive outcome (cooperate, cooperate).
This ambiguity raises critical questions: What determines the final state of the system? Are there underlying conditions on the game structure that guarantee convergence to the competitive equilibrium and thus prevent collusion? 
Answering our central question is clearly not a simple ``yes" or ``no.'' This motivating example demands a deeper theoretical investigation into the dynamics of interacting Thompson sampling agents, which is the primary objective of this paper.

\subsection{Our Contributions}
This paper provides a definitive theoretical resolution to the puzzle presented by our motivating experiment. 
Our central contribution is the identification of a set of sufficient conditions on the game's payoff matrix that guarantee interacting Thompson sampling agents will converge to the unique pure-strategy Nash equilibrium, thereby precluding the possibility of algorithmic collusion. This result provides, for the first time, a clear criterion to distinguish between games where Thompson sampling is ``safe'' from a competitive standpoint and those where it may be susceptible to collusive outcomes.

To establish this result, we make the following technical and conceptual contributions:

\paragraph{A Stochastic Approximation Model for Interacting Thompson Sampling.}
We begin by formalizing the learning dynamics under a standard and realistic information structure where players only observe their own actions and resulting payoffs, without access to their opponent's choices or rewards. We demonstrate that the evolution of the players' posterior beliefs (means and variances) can be precisely modeled as a stochastic process. This formulation places the problem within the powerful framework of stochastic approximation (SA) \citep{kushner2003stochastic}, allowing us to analyze the system's long-run behavior.

\paragraph{A Novel Proof Technique for Highly Asynchronous SA Systems.}
Our primary technical innovation is the development of a new analytical method to prove the convergence of this system. A direct application of existing SA theory fails for two fundamental reasons inherent to Thompson sampling dynamics:
\begin{itemize}
	\item \textbf{Asynchronicity:} As a player's algorithm begins to favor a particular action, the suboptimal actions are chosen with vanishing frequency (on the order of $O(\log n)$ times in $n$ rounds). This means the posterior beliefs for different actions are updated on vastly different time scales ($O(n)$ vs. $O(\log n)$), violating the assumptions of standard asynchronous SA theories which require update frequencies to be of the same order.
	\item \textbf{Lack of Global Lipschitz Continuity:} The vector field that drives the system's evolution is not globally Lipschitz continuous. Specifically, when the posterior means of two actions are nearly equal, a small perturbation can dramatically change the probability of selecting either action, creating a ``sharp" transition in the dynamics that violates a key assumption for many convergence proofs in SA.
\end{itemize}

To overcome these obstacles, we develop a novel, sample-path-wise argument from first principles. While inspired by the spirit of \cite{tsitsiklis1994asynchronous}, our approach is substantially extended to handle the extreme asynchronicity and the specific non-Lipschitz nature of our system. This proof strategy is, to our knowledge, new to the literature and may be of independent interest for analyzing other learning algorithms with similar dynamic properties. The relationship of our analytical approach to existing SA literature is summarized in Table~\ref{tab:literature}.

\begin{table}[h]
	\caption{Connection to the literature on proving the convergence of the system}
	\label{tab:literature}
	\begin{tabular}{lll}
		\toprule 
		Approach & Literature & Challenge\\
		\midrule
		SA with two time scales &\cite{borkar1997stochastic, borkar2009stochastic}& Not updated simultaneously\\
		Asynchronous SA &\cite{borkar1998asynchronous}& Updated with very different frequencies\\
		Sample-path-wise argument &\cite{tsitsiklis1994asynchronous} &Not globally Lipschitz continuous \\
		\bottomrule
	\end{tabular}
\end{table}

\subsection{Literature Review}\label{sec:literature}
Algorithmic collusion has attracted the attention of scholars and regulators recently.
\cite{calvano2020artificial} demonstrate using simulation that when two competing firms both use Q-learning algorithms, the set prices may converge to an collusive equilibrium higher than the Nash equilibrium, although the two firms do not collude explicitly.
Similar phenomena have been observed for UCB \citep{hansen2021frontiers} or more sophisticated learning algorithms \citep{meylahn2022learning,aouad2021algorithmic,klein2021autonomous}.
Empirically, \cite{brown2023competition} use high-frequency online retail data to find that some retailers deploy algorithms that react faster or commit to prices for varying durations.
\cite{banchio2022artificial} use a theoretical framework to explain why independent learning agents using adaptive algorithms might coordinate their behavior tacitly.
\cite{wang2025contextual} argue that the presence of contextual information may lead to a supracompetitive outcome under a predictive algorithm, even when the algorithm itself doesn't lead to collusion. 
\cite{keppo2025ai} analyze how heterogeneity in fundamental features of LLM-based pricing agents and in external market conditions will affect algorithmic collusion theoretically and experimentally.
In this study, we prove that algorithmic collusion will not arise for Thompson sampling under a mild condition.
This is surprising given that many other popular learning algorithms will lead to algorithmic collusion.

Repeated games and learning have been a classic topic in economics \citep{fudenberg1998theory}. 
Fictitious play is a simple learning rule where players assume that opponents will use similar strategies as in the past and respond optimally to the historical empirical distribution, which has been studied extensively \citep{robinson1951iterative, miyasawa1963convergence, krishna1998convergence, hofbauer2002global}. 
For other algorithms, it has been shown that if they achieve sublinear regret \citep{cesa2006prediction}, then in the case that all players use such an algorithm, the time-average behavior will converge to the set of coarse correlated equilibria (CCE), also known as the game’s Hannan set.
While average payoffs may converge, the strategies themselves might not \citep{mertikopoulos2018cycles}. This motivates the ongoing search for last-iterate guarantees, which are theoretically stronger convergence results and can describe the evolution of the players action profiles, such as \cite{mertikopoulos2017convergence,mertikopoulos2019learning,mazumdar2020gradient,cai2023uncoupled}.
Achieving last-iterate convergence can be challenging, and it often requires specific algorithmic designs or assumptions on the game structure, such as the optimistic gradient \citep{golowich2020tight}, extra-gradient \citep{golowich2020last} and monotone games \citep{lin2020finite, cai2022finite, ba2025doubly}. 
This study also focuses on the last-iterate convergence.
The main difference of our setup is that the actions are not continuous and the players do not receive first-order feedback.
This setup is first proposed in \cite{ortega2014generalized} and the convergence is shown numerically.
\cite{pmlr-v161-o-donoghue21a} show that using Thompson sampling in matrix games with bandit feedback when the competitor plays a different policy can lead to linear regret. In contrast, in our setup, both players use Thompson sampling and do not observe the actions of their opponents.

Our study deviates from multi-agent reinforcement learning in terms of the motivation and research question.
In multi-agent reinforcement learning \citep{zhang2021multi,yang2018mean,cai2023uncoupled} or multi-agent Thompson sampling \citep{verstraeten2020multi}, 
a typical goal is to design algorithms and communication protocols that only rely on the local information of each agent to achieve convergence to the cooperative or Nash equilibrium.
In our study, we do not design new algorithms but document the dynamics under the classic Thompson sampling.
There is no communication between the players either.
Thompson sampling has been a popular algorithm for stochastic Multi-Armed Bandit (MAB, \cite{lattimore2020bandit}). A tutorial of Thompson sampling is given in \cite{russo2018tutorial} and the theoretical property is proved in, e.g., \cite{kaufmann2012thompson,agrawal2017near}.

There is also a line of research on independent learning in multi-agent settings, see e.g. \citep{ozdaglar2021independent, ding2022independent, chen2023finite}, which is closely related to our work. In this paradigm, individual agents neglect the complex interactions between them and
treat other agents as part of a non-stationary environment due to the actions
taken by other agents. However, independent learning generally lacks convergence guarantees \citep{zhang2021multi}, except in special cases such as zero-sum stochastic games \citep{chen2023finite} or Markov potential games \citep{ding2022independent}. Our research addresses this gap by investigating the convergence properties of independent learning with Thompson Sampling in general-sum matrix games. Moreover, we analyze the popular TS algorithm to study the potential for collusion, instead of best-response type algorithms often used in independent learning. 
Therefore, our motivation and technical analysis are significantly different from existing studies on independent learning in games.

Finally, we mention a growing body of literature recently on finite-time analysis of SA, see, e.g., \cite{srikant2019finite, qu2020finite, chen2021lyapunov, haque2023tight, durmus2025finite} and references therein.
Although we focus on the asymptotic analysis, which is a weaker notion than a finite-time analysis,
none of these recent approaches/results on finite time analysis can be applied to our problem. Note that the recent advance is to push the existing asymptotic analysis to the non-asymptotic regime, while the challenge arising from our setting is regarding the asymptotic analysis itself. For example, \cite{haque2023tight} study finite-time analysis of two-time scale SA, but our SA scheme is not a two-time scale SA (because the posterior of the inferior action is only updated infrequently and sporadically). Similarly, \cite{qu2020finite} study finite-time analysis of asynchronous SA, but our SA scheme is also different from their asynchronous SA, because the updating frequencies of different actions are not of the same order. Therefore, we can not apply these recent approaches in the recent SA literature. Our asymptotic convergence analysis of the SA scheme that arises in the game setting with Thompson sampling is novel, and this is one of our main technical contributions.


\section{Two-Player Game with Thompson Sampling}
\subsection{Problem Formulation}\label{sec:formulation}
Consider a game with two players, Player 1 and Player 2, having sets of actions $\Iscr \coloneqq \{1,\dots,I\}$ and $\Jscr \coloneqq  \{1,\dots,J\}$ respectively. 
The payoff of the game is represented by $G=(A,B)$,
where $A,B \in \R^{I \times J}$ are the expected payoff matrices.
In particular, when actions $i\in \Iscr$ and $j\in \Jscr$ are taken by Player 1 and 2, respectively, 
the expected payoffs of Player 1 and 2 are $A_{i,j}$ and $B_{i,j}$, respectively.
The game is played repeatedly.
We use $i_n$ and $j_n$ to denote the actions taken by Player 1 and Player 2 in round $n$.
Given $i_n$ and $j_n$, the \emph{realized} and \emph{observed} payoffs in round $n$ for Player 1 and 2 are (samples of) normal random variables respectively:
$a_{i_n,n} \sim \mathcal{N}(A_{i_n,j_n},1)$ and $b_{j_n,n} \sim \mathcal{N}(B_{i_n,j_n},1)$,
where $\mathcal{N}(\mu,\sigma^2)$ denotes the normal distribution with mean $\mu$ and variance $\sigma^2$. We assume the variance of the observed payoffs is fixed at 1 for all actions and players.

Both players employ Thompson sampling, an algorithm originally designed for single-agent MAB problems \citep{russo2018tutorial}. In this setting, each player treats the repeated game as a personal MAB problem, where their actions are the ``arms.'' Crucially, each player's algorithm operates independently, without knowledge of the other player's existence, actions, or payoffs. This setup represents an uncoupled learning dynamic.

The Thompson sampling procedure for each player is as follows. Each player maintains a posterior distribution for the expected payoff of each of their actions. We assume a Gaussian likelihood for the observed payoffs and a conjugate Gaussian prior for the unknown expected payoffs. Consequently, the posterior distribution for each action's expected payoff is also Gaussian. In each round $n$, a player first samples a random value for each action from its current posterior distribution. Then, the player selects the action corresponding to the highest sampled value. After observing the realized payoff from the chosen action, the player updates the posterior distribution for that action using Bayes' rule. This process is formally described in Algorithm~\ref{alg:ts1}.

\begin{algorithm}[h!]
	\caption{A Two-Player Game with Thompson Sampling}
	\label{alg:ts1}
	\begin{algorithmic}[1]
		\Require Prior means $x_{i,0}$ for $i \in \Iscr$ and $y_{j,0}$ for $j \in \Jscr$. Prior variance is 1 for all actions.
		\State \textbf{Initialize:} 
		\State For Player 1 (for all $i \in \Iscr$): number of pulls $N_{i,0} \leftarrow 0$; sum of payoffs $R_{i,0} \leftarrow x_{i,0}$.
		\State For Player 2 (for all $j \in \Jscr$): number of pulls $M_{j,0} \leftarrow 0$; sum of payoffs $T_{j,0} \leftarrow y_{j,0}$.
		\For{$n=1,2,\dots$}
		\State \textbf{Player 1's Action Selection:}
		\State \quad For each action $k \in \Iscr$, sample $\theta_{k,n} \sim \mathcal{N}\left(\frac{R_{k,n-1}}{N_{k,n-1}+1}, \frac{1}{N_{k,n-1}+1}\right)$.
		\State \quad Choose action $i_n = \argmax_{k \in \Iscr} \theta_{k,n}$.
		\State \textbf{Player 2's Action Selection:}
		\State \quad For each action $k \in \Jscr$, sample $\vartheta_{k,n} \sim \mathcal{N}\left(\frac{T_{k,n-1}}{M_{k,n-1}+1}, \frac{1}{M_{k,n-1}+1}\right)$.
		\State \quad Choose action $j_n = \argmax_{k \in \Jscr} \vartheta_{k,n}$.
		\State \textbf{Observe Payoffs:} Players observe $a_{i_n,n} \sim \mathcal{N}(A_{i_n,j_n},1)$ and $b_{j_n,n} \sim \mathcal{N}(B_{i_n,j_n},1)$.
		\State \textbf{Update Statistics for Player 1:}
		\State \quad $N_{i_n,n} \leftarrow N_{i_n,n-1} + 1$ and $R_{i_n,n} \leftarrow R_{i_n,n-1} + a_{i_n,n}$.
		\State \quad For $k \ne i_n$, set $N_{k,n} \leftarrow N_{k,n-1}$ and $R_{k,n} \leftarrow R_{k,n-1}$.
		\State \textbf{Update Statistics for Player 2:}
		\State \quad $M_{j_n,n} \leftarrow M_{j_n,n-1} + 1$ and $T_{j_n,n} \leftarrow T_{j_n,n-1} + b_{j_n,n}$.
		\State \quad For $k \ne j_n$, set $M_{k,n} \leftarrow M_{k,n-1}$ and $T_{k,n} \leftarrow T_{k,n-1}$.
		\EndFor
	\end{algorithmic}
\end{algorithm}
In our model, players operate with limited information: they only observe their own historical actions and resulting payoffs, without any knowledge of their competitor's choices or outcomes. This information structure reflects many real-world business environments where competitive data is scarce or confidential. It is a particularly relevant setting for studying algorithmic collusion, as regulators are concerned about market outcomes when firms independently use data-driven algorithms. In such a scenario, it is plausible for a firm to deploy a standard single-agent learning algorithm to optimize its decisions. We focus on Thompson sampling due to its widespread use and strong performance in the single-agent context.

A central challenge in this analysis is that the Thompson sampling algorithm is \emph{not correctly specified} for this game environment. From each player's perspective, the opponent's actions introduce non-stationarity into the observed payoffs, violating a key assumption of standard MAB analysis. Furthermore, although the players' algorithms are uncoupled, their actions become intricately linked through the game's payoff structure. The payoff a player observes depends on the opponent's action, which in turn influences the opponent's future posterior beliefs and actions. This creates a complex, nonlinear feedback loop. A priori, it is unclear whether this dynamic system will stabilize or converge. This paper demonstrates that, perhaps surprisingly, the players' joint actions converge almost surely to the pure-strategy Nash equilibrium under a set of conditions. Consequently, our results suggest that algorithmic collusion does not arise when independent firms use Thompson sampling in this setting.

\subsection{Game Dynamics} \label{sec:state-dynamics}
To analyze the game's evolution, we define the state variables that capture the players' posterior beliefs at each round. These variables are updated based on the history of actions and observed payoffs.

For Player 1, recall that $N_{i,n} \coloneqq \sum_{s=1}^n \1_{\{i_s=i\}}$ is the number of times action $i \in \Iscr$ has been chosen up to round $n$. We define the following state variables for Player 1:
\begin{align}\label{eq:x-w} 
	x_{i,n} &\coloneqq \frac{x_{i,0}+\sum_{s=1}^n a_{i_s,s} \cdot \1_{\{i_s=i\}}}{N_{i,n}+1},  \\
	w_{i,n} &\coloneqq \frac{1}{N_{i,n}+1}. \notag
\end{align}
As is standard for Bayesian inference with a Gaussian likelihood and conjugate prior, $x_{i,n}$ and $w_{i,n}$ represent the mean and variance, respectively, of the posterior Gaussian distribution for the expected payoff of action $i$ at the end of round $n$ (see, e.g., \cite{russo2018tutorial}). The initial state at $n=0$ corresponds to the prior distribution, with $N_{i,0}=0$, posterior mean $x_{i,0}$, and posterior variance $w_{i,0}=1$.

The state variables for Player 2 are defined symmetrically. For each action $j \in \Jscr$, let $M_{j,n} \coloneqq \sum_{s=1}^n \1_{\{j_s=j\}}$ be the number of times action $j$ has been chosen. The posterior mean and variance are:
\begin{align} 
	y_{j,n} &\coloneqq \frac{y_{j,0}+\sum_{s=1}^n b_{j_s,s} \cdot \1_{\{j_s=j\}}}{M_{j,n}+1}, \label{eq:y_def} \\
	v_{j,n} &\coloneqq \frac{1}{M_{j,n}+1}. \label{eq:v_def}
\end{align}

The complete state of the system at the end of round $n$ is captured by the vector $\bm S_n \in \R^{I+J} \times (0,1]^{I+J}$, defined as:
\begin{align}\label{def:Sn}
	\bm S_n \coloneqq \big(x_{1,n},\dots, x_{I,n}, y_{1,n},\dots,y_{J,n}, w_{1,n},\dots,w_{I,n},v_{1,n},\dots,v_{J,n}\big).
\end{align}
Note that $\bm S_n$ is a sufficient statistic for the players' decision-making process in round $n+1$, as it fully specifies the posterior distributions required by Algorithm~\ref{alg:ts1}.

We now analyze the dynamics of the state vector $\bm S_n$. Our goal is to express its evolution in the form of a stochastic approximation (SA) algorithm, which will allow us to study its asymptotic behavior. The general form of an SA update is $\bm S_{n+1} = \bm S_n + \text{step\_size} \cdot (\text{mean\_field} - \bm S_n + \text{noise})$.

Let's first derive the exact update rule for Player 1's state variables, $(x_{i,n}, w_{i,n})$. The dynamics for Player 2 are symmetric. If Player 1 chooses action $i \in \Iscr$ in round $n+1$, so that $i_{n+1}=i$, the number of pulls becomes $N_{i,n+1} = N_{i,n}+1$. The update rules for the posterior mean and variance can be written as:
\begin{align}
	x_{i,n+1} - x_{i,n} &= \frac{1}{N_{i,n+1}+1} (a_{i,n+1} - x_{i,n}) = w_{i,n+1} (a_{i,n+1} - x_{i,n}), \\		
	w_{i,n+1} - w_{i,n} &= \frac{1}{N_{i,n+1}+1} (-w_{i,n}) = w_{i,n+1} (-w_{i,n}).
\end{align}
If action $i$ is not chosen, the state variables for that action remain unchanged. We can combine these cases using an indicator function. Let $\alpha_{i,n+1} \coloneqq \1_{\{i_{n+1}=i\}} w_{i,n+1}$ be the state-dependent random step size for action $i$. The dynamics for all $i \in \Iscr$ are then:
\begin{align}
	x_{i,n+1} &= x_{i,n} + \alpha_{i,n+1} (a_{i,n+1} - x_{i,n}), \label{eq:x_update_compact} \\
	w_{i,n+1} &= w_{i,n} + \alpha_{i,n+1} (-w_{i,n}). \label{eq:w_update_compact}
\end{align}

To analyze these updates, we decompose them into a drift term (the conditional expectation given the current state $\bm S_n$) and a martingale difference noise term. This requires characterizing the distributions of the step size $\alpha_{i,n+1}$ and the reward $a_{i,n+1}$.

\paragraph{Choice Probabilities.}
The probability that Player 1 chooses action $i$ in round $n+1$, conditional on the history $\Fscr_n$ (which is summarized by $\bm S_n$), is
\begin{align}\label{eq:phi_def}
	\varphi_{i}(\bm S_n) \coloneqq \PR(i_{n+1}=i|\Fscr_n) = \PR(\theta_{i,n+1} > \max_{k \ne i}\theta_{k,n+1}|\bm S_n),
\end{align}
where each $\theta_{k,n+1} \sim \mathcal{N}(x_{k,n}, w_{k,n})$ is sampled independently. To express this probability in a closed form, we analyze the joint distribution of the differences between the sampled values. Let us define a vector of $I-1$ random variables $\bm Y^{(i)} \coloneqq (Y_k^{(i)})$ for $k \ne i$, where $Y_k^{(i)} \coloneqq \theta_{k,n+1} - \theta_{i,n+1}$. The event that action $i$ is chosen, ${\theta_{i,n+1} > \theta_{k,n+1} \text{ for all } k \ne i}$, is equivalent to the event ${Y_k^{(i)} < 0 \text{ for all } k \ne i}$.

Since each $\theta_{k,n+1}$ is an independent normal random variable, their linear combination, $\bm Y^{(i)}$, follows a multivariate normal distribution. We can find its parameters as follows. For any $k,l \in \Iscr \setminus \{i\}$:
\begin{itemize}
	\item \textbf{Mean:} $\E[Y_k^{(i)}] = \E[\theta_{k,n+1}] - \E[\theta_{i,n+1}] = x_{k,n} - x_{i,n}$.
	\item \textbf{Variance:} $\Var(Y_k^{(i)}) = \Var(\theta_{k,n+1}) + \Var(\theta_{i,n+1}) = w_{k,n} + w_{i,n}$.
	\item \textbf{Covariance ($k \ne l$):} $\Cov(Y_k^{(i)}, Y_l^{(i)}) = \Cov(\theta_{k,n+1} - \theta_{i,n+1}, \theta_{l,n+1} - \theta_{i,n+1}) = \Var(\theta_{i,n+1}) = w_{i,n}$, due to the independence of $\theta_{k,n+1}$, $\theta_{l,n+1}$, and $\theta_{i,n+1}$.
\end{itemize}
The choice probability is thus $\varphi_{i}(\bm S_n) = \PR(\bm Y^{(i)} < \bm 0)$. To express this using a standard normal CDF, we standardize the vector $\bm Y^{(i)}$. Let $\bm Z^{(i)}$ be the standardized version of $\bm Y^{(i)}$, where each component is $Z_k^{(i)} = (Y_k^{(i)} - \E[Y_k^{(i)}]) / \sqrt{\Var(Y_k^{(i)})}$. The vector $\bm Z^{(i)}$ has a zero mean and a correlation matrix $\bm \rho_{n}^{(i)}$. The condition $Y_k^{(i)} < 0$ is equivalent to
\begin{align*}
	Z_k^{(i)} < \frac{0 - \E[Y_k^{(i)}]}{\sqrt{\Var(Y_k^{(i)})}} = \frac{-(x_{k,n} - x_{i,n})}{\sqrt{w_{k,n} + w_{i,n}}} = \frac{x_{i,n} - x_{k,n}}{\sqrt{w_{i,n} + w_{k,n}}}.
\end{align*}
Let $c_{k,n}^{(i)} \coloneqq (x_{i,n}-x_{k,n})/\sqrt{w_{i,n}+w_{k,n}}$ for $k \ne i$. The probability $\varphi_{i}(\bm S_n)$ is the probability that $Z_k^{(i)} < c_{k,n}^{(i)}$ for all $k \ne i$. This is precisely the definition of the CDF of $\bm Z^{(i)}$ evaluated at the point $(c_{k,n}^{(i)})_{k\neq i}$. Therefore,
\begin{align}\label{notation:phi-mul}
	\varphi_{i,n+1}\coloneqq\varphi_{i}(\bm S_n) = \Phi_{I-1}\left((c_{k,n}^{(i)})_{k \ne i}; \bm \rho_{n}^{(i)}\right),
\end{align}
where $\Phi_{I-1}$ is the CDF of a $(I-1)$-variate standard normal distribution with correlation matrix $\bm \rho_{n}^{(i)}$. The $(k,l)$-th off-diagonal entry of this matrix is given by
\begin{align*}
	(\bm \rho_{n}^{(i)})_{k,l} = \frac{\Cov(Y_k^{(i)}, Y_l^{(i)})}{\sqrt{\Var(Y_k^{(i)})\Var(Y_l^{(i)})}} = \frac{w_{i,n}}{\sqrt{(w_{i,n}+w_{k,n})(w_{i,n}+w_{l,n})}}.
\end{align*}
Because the correlation coefficients are all positive,
by Slepian's inequality \citep{joag1983association}, we have the following lower bound which will be used in our convergence analysis:
\begin{align}\label{inequ:phi-mul}
	\varphi_{i}(\bm S_n) \geq \prod_{k \ne i} \Phi\left( c_{k,n}^{(i)} \right),
\end{align}
where $\Phi(\cdot)$ is the standard normal CDF. Symmetrically, we define Player 2's choice probability for action $j \in \Jscr$ as $\psi_{j}(\bm S_n)$.
The details of the derivation can be found in Appendix \ref{sec:prob}.

\paragraph{Reward Decomposition.}
The reward $a_{i,n+1}$ observed by Player 1, when selecting action $i$ in round $n+1$, depends on Player 2's action $j_{n+1}$. We can write $a_{i,n+1} = A_{i,j_{n+1}} + \epsilon_{n+1}$, where $\epsilon_{n+1} \sim \mathcal{N}(0,1)$. To isolate the martingale noise, we condition on $\Fscr_n$: 
\begin{align}
	\E[a_{i,n+1} | \Fscr_n] = \sum_{j \in \Jscr} A_{i,j} \PR(j_{n+1}=j|\Fscr_n) = \sum_{j \in \Jscr} A_{i,j} \psi_{j}(\bm S_n).
\end{align}
Thus, we can decompose the reward as $a_{i,n+1} = \sum_{j \in \Jscr} A_{i,j} \psi_{j}(\bm S_n) + \bar{a}_{i,n+1}$, where the noise term
\begin{align}\label{bar-a}
	\bar{a}_{i,n+1} \coloneqq \left(\sum_{j \in \Jscr} A_{i,j} \1_{\{j_{n+1}=j\}} - \sum_{j \in \Jscr} A_{i,j} \psi_{j}(\bm S_n)\right)  + \epsilon_{n+1}
\end{align}
is a martingale difference sequence, i.e., $\E[\bar{a}_{i,n+1} | \Fscr_n] = 0$. A similar term $\bar{b}_{j,n+1}$ exists for Player 2.

\paragraph{Stochastic Approximation Form.}
Substituting the decomposed reward into \eqref{eq:x_update_compact} and defining analogous step sizes $\beta_{j,n+1} \coloneqq \1_{\{j_{n+1}=j\}} v_{j,n+1}$ for Player 2, we obtain the full set of dynamic equations:
\begin{align}
	x_{i,n+1} &= x_{i,n} + \alpha_{i,n+1} \left(\sum_{j \in \Jscr} A_{i,j} \psi_{j}(\bm S_n) - x_{i,n} + \bar{a}_{i,n+1}\right), \\
	y_{j,n+1} &= y_{j,n} + \beta_{j,n+1} \left(\sum_{i \in \Iscr} B_{i,j} \varphi_{i}(\bm S_n) - y_{j,n} + \bar{b}_{j,n+1}\right), \\ 
	w_{i,n+1} &= w_{i,n} + \alpha_{i,n+1} (-w_{i,n}), \\
	v_{j,n+1} &= v_{j,n} + \beta_{j,n+1} (-v_{j,n}).
\end{align}
We can now write this system in a compact vector form. Let the mean field function $F: \R^{I+J} \times (0,1]^{I+J} \to \R^{2(I+J)}$ be
\begin{align}\label{eq:F-mul}
	F(\bm S) \coloneqq \left( \left(\sum_{j \in \Jscr} A_{i,j} \psi_{j}(\bm S)\right)_{i \in \Iscr}, \left(\sum_{i \in \Iscr} B_{i,j} \varphi_{i}(\bm S)\right)_{j \in \Jscr}, \bm 0_{I+J} \right).
\end{align}	
The dynamics of the state vector $\bm S_n$ from \eqref{def:Sn} can then be expressed as:
\begin{align}\label{eq:state-dynamics}
	\bm S_{n+1} - \bm S_{n} = \bm \gamma_{n+1} \circ \left(F(\bm S_{n}) - \bm S_{n} + \bar{\bm{\xi}}_{n+1}\right),
\end{align}
where $\circ$ denotes element-wise multiplication. The random step-size vector is $\bm \gamma_{n+1} \coloneqq \\ \left( \left(\alpha_{i,n+1}\right)_{i \in \Iscr}, \left(\beta_{j,n+1}\right)_{j \in \Jscr}, \left(\alpha_{i,n+1}\right)_{i \in \Iscr}, \left(\beta_{j,n+1}\right)_{j \in \Jscr}\right) \in (0,1]^{2(I+J)}$. The noise vector $\bar{\bm{\xi}}_{n+1}$ contains the martingale difference terms $\bar{a}_{i,n+1}$ and $\bar{b}_{j,n+1}$ in the appropriate positions and zeros elsewhere.

{Equation \eqref{eq:state-dynamics} can be viewed a Stochastic Approximation (SA) update. Unlike standard SA schemes, the step size in \eqref{eq:state-dynamics} is random, state-dependent, and does not necessarily decrease due to the indicator functions involved in $\alpha_{i,n}$ and $\beta_{j,n}$.
	The term $F(\bm S_n) - \bm S_n$ represents the expected direction of motion, or drift, while $\bar{\bm{\xi}}_{n+1}$ is the zero-mean noise. This formulation is the foundation for our convergence analysis. }

\subsection{Assumptions and System Equilibrium} \label{sec:equi-game}
Having cast the system dynamics in the stochastic approximation form \eqref{eq:state-dynamics}, we now identify the equilibrium point of its associated mean field dynamics, $F(\bm S) - \bm S$. The stability of this point is central to our analysis, and for it to be well-defined, we impose a standard assumption on the underlying game.

\begin{assumption}\label{asp:unique-ne}
	The game satisfies two conditions:
	\begin{enumerate}
		\item (No ties) For any player, the payoffs for different actions are distinct given any action of the opponent. That is, $A_{i,j} \neq A_{i',j}$ for all $j$ and $i \neq i'$, and $B_{i,j} \neq B_{i,j'}$ for all $i$ and $j \neq j'$.
		\item (Unique Pure NE) The game possesses a unique pure-strategy Nash equilibrium (NE).
	\end{enumerate}
\end{assumption}

The no-ties condition is a mild, technical requirement common in the learning-in-games literature (e.g., \cite{cesa2006prediction, wunder2010classes}). The assumption of a unique pure-strategy NE is more substantial and ensures that the learning dynamics have a single, unambiguous target. This is a common setup for proving convergence in games with continuous action spaces or complex dynamics \citep{jordan2025adaptive}. While games can have mixed-strategy equilibria, our numerical results in Section~\ref{sec:experiments} suggest that in such cases, the Thompson sampling dynamics may not converge or may converge to one of the pure-strategy profiles, motivating our focus on the unique pure NE case for analytical tractability.

Without loss of generality, let the action profile $(1,1)$ be the unique pure-strategy NE. This implies $A_{1,1} > A_{i,1}$ for all $i \neq 1$ and $B_{1,1} > B_{1,j}$ for all $j \neq 1$. Under this assumption, the mean field dynamics have a unique equilibrium point $\bm S^*$ defined as:
\begin{align} 
	\bm S^* &\coloneqq (x_1^*, \dots, x_I^*, y_1^*, \dots, y_J^*, w_1^*,\dots, w_I^*, v_1^*, \dots, v_J^*) \nonumber \\
	&= \big( (A_{i,1})_{i \in \Iscr}, (B_{1,j})_{j \in \Jscr}, \bm{0}_{I+J} \big). \label{def:S-star-mul}
\end{align}
To see that $\bm S^*$ is indeed an equilibrium point, we verify that $F(\bm S^*) - \bm S^* = \bm 0$. At $\bm S = \bm S^*$, the posterior variances are all zero ($w_i^*=0, v_j^*=0$). For Player 1, the posterior mean for action 1 is $x_1^* = A_{1,1}$, while for any other action $i \neq 1$, $x_i^* = A_{i,1}$. Since $A_{1,1} > A_{i,1}$, the choice probability $\varphi_1(\bm S^*) = 1$ and $\varphi_i(\bm S^*) = 0$ for $i \neq 1$. Symmetrically, $\psi_1(\bm S^*) = 1$ and $\psi_j(\bm S^*) = 0$ for $j \neq 1$. Plugging these into the definition of $F(\bm S^*)$ in \eqref{eq:F-mul}, the first $I$ components become $(\sum_{j} A_{i,j} \psi_j(\bm S^*))_{i \in \Iscr} = (A_{i,1})_{i \in \Iscr}$, which are precisely the $x^*$ components of $\bm S^*$. A similar calculation holds for the $y^*$ components, and the variance components are zero by definition. Thus, $F(\bm S^*) = \bm S^*$.

The main theoretical contribution of this paper is to prove that the state sequence $\{\bm S_n\}$ generated by the Thompson sampling dynamics converges almost surely to this equilibrium point $\bm S^*$.

\section{Main Results}\label{sec:results}
\subsection{Preliminary Results}
Our proof of convergence for the state vector $\bm S_n$ to the equilibrium $\bm S^*$ relies on the theory of stochastic approximation. This requires verifying a set of foundational conditions on the step sizes, noise process, and boundedness of the iterates. While these conditions are standard in form, proving them in our strategic learning context is non-trivial due to the coupling between the players' actions. The proofs for the following lemmas are detailed in the Appendix.

The first critical step is to ensure that learning never ceases for any action.
\begin{lemma}[Infinite Exploration]\label{lemma:infinite}
	Almost surely, every action is played infinitely often by both players. That is, $\lim_{n \to \infty} N_{i,n}=\infty$ for all $i \in \Iscr$, and $\lim_{n \to \infty} M_{j,n}=\infty$ for all $j \in \Jscr$.
\end{lemma}

This infinite exploration is essential for the step sizes to satisfy the classic Robbins-Monro conditions, which are crucial for convergence. Recall the \textit{random} step sizes $\alpha_{i,n} \coloneqq \1_{\{i_n=i\}} w_{i,n}$ and $\beta_{j,n} \coloneqq \1_{\{j_n=j\}} v_{j,n}$.

\begin{lemma}[Step-Size Conditions]\label{lem:step_size}
	The step-size sequences for each action satisfy, almost surely,
	\begin{enumerate}
		\item $\sum_{n=1}^{\infty} \alpha_{i,n} = \infty$ and $\sum_{n=1}^{\infty} \beta_{j,n} = \infty$ for all $i \in \Iscr, j \in \Jscr$.
		\item $\sum_{n=1}^{\infty} (\alpha_{i,n})^2 < \infty$ and $\sum_{n=1}^{\infty} (\beta_{j,n})^2 < \infty$ for all $i \in \Iscr, j \in \Jscr$.
	\end{enumerate}
\end{lemma}

Lemma~\ref{lem:step_size} is a cornerstone of the analysis. The first condition ensures that the accumulated step sizes are sufficient to overcome any initial state and reach the equilibrium. The second condition guarantees that the step sizes diminish quickly enough to quell the effects of noise, allowing the iterates to stabilize. The primary challenge here, distinct from standard multi-armed bandit or SA analysis, is to show that one player's strategy does not inadvertently confound an opponent's action, which would violate the first condition. Our proof confirms that the exploration inherent in Thompson sampling is robust to this strategic interaction.

Next, we establish that the noise term in our SA formulation is well-behaved and that the process remains stable.
Define
\begin{align}\label{eq:noise-mul}
	\bar{\bm{\xi}}_{n+1}:=(\bar{a}_{1,n+1}, \dots, \bar{a}_{I,n+1}, \bar{b}_{1,n+1}, \dots, \bar{b}_{J,n+1}, 0,\dots,0),
\end{align}
where $\bar{a}_{i,n+1}$ is given in \eqref{bar-a} and $\bar{b}_{j,n+1}$ can be defined similarly. 
\begin{lemma}[Martingale Difference Noise]\label{lem:martingale_diff} 
	The noise sequence $(\bar{\bm{\xi}}_{n}: n \ge 1)$ defined in \eqref{eq:noise-mul} is a martingale difference sequence with respect to the filtration $\Fscr_{n-1}$, satisfying $\E[\bar{\bm \xi}_{n}|\mathcal{F}_{n-1}]=0$. Furthermore, its conditional second moment is uniformly bounded, i.e., there exists a constant $C < \infty$ such that $\E[\|\bar{\bm \xi}_{n}\|^2|\Fscr_{n-1}] \leq C$ for all $n$.
\end{lemma}

\begin{lemma}[Boundedness of Iterates]\label{lem:bounded}
	The sequence of state vectors $\{\bm S_n\}$ is bounded almost surely, i.e., $\sup_{n} \| \bm S_n \| < \infty$.
\end{lemma}

These final two lemmas provide the necessary stability properties. Lemma~\ref{lem:martingale_diff} confirms that the noise does not systematically bias the updates, while Lemma~\ref{lem:bounded} ensures that the state estimates do not diverge, confining the dynamics to a compact set where convergence can be analyzed.

Together, these four lemmas provide the essential scaffolding for our main convergence theorem. However, because the step sizes are state-dependent and jointly determined by both players' learning dynamics, the classical convergence proofs for SA must be adapted. We address this in the subsequent section.

\subsection{Main Results}\label{sec:thm-analysis}
We now present the main result of our analysis: the almost sure convergence of the learning dynamics to the state corresponding to the game's unique pure-strategy Nash Equilibrium. As before, we assume without loss of generality that $(1,1)$ is the unique pure NE.

The convergence proof requires a technical condition on the payoff matrices. This condition ensures that the payoff differences at the NE strategy profile are sufficiently dominant to prevent the players' beliefs from lingering in regions of high uncertainty.

\begin{assumption}\label{asp:payoff-matrix-mul}
	The payoff matrices $(A,B)$ satisfy for all distinct pairs $i, \ell \in \Iscr$ and $j, k \in \Jscr$:
	\begin{align}
		\max_{m \ne 1} |A_{i,1}-A_{i,m}| + \max_{m \ne 1} |A_{\ell,1}-A_{\ell,m}| &< |A_{i,1}-A_{\ell,1}|, \\
		\max_{m \ne 1} |B_{1,j}-B_{m,j}| + \max_{m \ne 1} |B_{1,k}-B_{m,k}| &< |B_{1,j}-B_{1,k}|.
	\end{align}
\end{assumption}
\noindent This assumption requires that for any two actions, the difference in their expected payoffs when the opponent plays the NE action (the right-hand side) is larger than the sum of the maximum possible payoff changes these actions could experience if the opponent were to deviate from their NE action (the left-hand side). It essentially imposes a form of stability on the NE payoffs. Our numerical results in Section~\ref{sec:experiments} suggest that the learning dynamics may not converge to the NE almost surely when Assumption~\ref{asp:payoff-matrix-mul} is voliated. 
We are now ready to state our main theorem.

\begin{theorem}\label{thm:main-result}
	Suppose Assumptions \ref{asp:unique-ne} (unique pure NE) and \ref{asp:payoff-matrix-mul} (payoff stability) hold. Then the state vector $\bm S_n$ defined in \eqref{def:Sn} converges almost surely to the equilibrium point $\bm S^*$ as $n\to\infty$, where $\bm S^*$ is defined in \eqref{def:S-star-mul}.
\end{theorem}

Theorem~\ref{thm:main-result} establishes the convergence of the players' posterior beliefs. Specifically, for Player 1, $x_{i,n} \to A_{i,1}$ and $w_{i,n} \to 0$ for all $i \in \Iscr$. This means Player 1 learns the true expected payoffs of all its actions under the condition that Player 2 plays its equilibrium action, 1. A symmetric result holds for Player 2.

A crucial consequence is that the players' action selections converge to the Nash Equilibrium. Since $(1,1)$ is the unique pure NE, we have $A_{1,1} > A_{k,1}$ for all $k \ne 1$. The convergence of beliefs thus implies that for large $n$, $x_{1,n} > x_{k,n}$ and the variances $w_{1,n}, w_{k,n}$ are small. Consequently, the term $(x_{1,n}-x_{k,n})/\sqrt{w_{1,n}+w_{k,n}}$ tends to $+\infty$. From the lower bound in \eqref{inequ:phi-mul}, it follows that the probability of Player 1 choosing the NE action converges to one:
\begin{align}\label{eq:lim-phi-mul}
	\lim_{n \to \infty} \varphi_{1}(\bm S_n) = 1, \quad \text{and} \quad \lim_{n \to \infty} \varphi_{k}(\bm S_n) = 0 \quad \text{for } k \ne 1.
\end{align}
Similarly, for Player 2, $\lim_{n \to \infty} \psi_{1}(\bm S_n) = 1$. This demonstrates that the joint action profile $(i_n, j_n)$ converges in probability to the NE $(1,1)$. This is a form of last-iterate convergence, which is a stronger guarantee than the convergence of the empirical frequency of play. 

The proof of Theorem \ref{thm:main-result} adapts the convergence analysis for asynchronous stochastic approximation in \cite{tsitsiklis1994asynchronous}. However, a direct application of existing theorems is not possible due to a key technical challenge: {the mean field mapping $F(\cdot)$ in \eqref{eq:F-mul} is not a contraction over the entire state space, because the choice probabilities $\varphi_i$ and $\psi_j$ are not even globally Lipschitz continuous.} 
Specifically, the gradient of $\varphi_i$ with respect to $x_k$ for $k \ne i$ can be shown to be proportional to 
\begin{align}
	\exp \left[-\frac{1}{2}\left(\frac{x_i-x_{k}}{\sqrt{w_{i}+w_{k}}}\right)^2 \right]\frac{1}{\sqrt{w_{i}+w_{k}}},
\end{align}
which explodes as beliefs become confident ($w_i, w_k \to 0$) but remain close ($x_i \approx x_k$). 

Our proof strategy is to show that the iterates $\bm S_n$ eventually avoid these ``singular'' regions. This is precisely the role of Assumption \ref{asp:payoff-matrix-mul}. It guarantees that the equilibrium point $\bm S^*$ is strictly separated from any state where $x_i=x_k$ or $y_j=y_k$. The core of the proof involves a sample-path argument showing that, for any trajectory, $\bm S_n$ eventually enters and remains within a compact set where the mapping $F(\cdot)$ is well-behaved and acts as a pseudo-contraction under the $\ell_{\infty}$ norm, which is sufficient to drive the iterates to the unique equilibrium $\bm S^*$.

\section{Proof Sketch of Theorem \ref{thm:main-result}}\label{sec:proof-sketch-thm}
The proof proceeds via a sample-path argument, analyzing the trajectory of the state vector $\{\bm S_n\}$ for a fixed realization of the random outcomes. The complete proof is long and hence deferred to Appendix~\ref{sec:proof-thm-mainresult}. Our strategy is tailored to overcome the central challenge that the mean field map $F(\cdot)$ is not globally a contraction, which precludes a direct application of standard stochastic approximation (SA) convergence theorems. The proof is structured in three main steps.

\paragraph{Step 1: Confining the Iterates to a Well-Behaved Region.}
The first and most critical step is to show that, for any sample path, the iterates $\bm S_n$ will almost surely eventually enter and remain in a region of the state space where the problematic non-Lipschitz behavior of $F(\cdot)$ does not occur.

To do this, we analyze the evolution of the error term $\bm S_n - \bm S^*$. The SA dynamics in \eqref{eq:state-dynamics} can be expanded into a component-wise recursive form for each state variable $S_{k,n}$ (see Lemma~\ref{lemma:Sn_recursion} in the Appendix):
\begin{align}\label{eq:three-terms}
	S_{k,n}-S_k^* =& \underbrace{(S_{k,0}-S_k^*) \prod_{\tau=1}^{n} (1-\gamma_{k,\tau})}_{C_{k,n}}
	+ \underbrace{\sum_{\tau=1}^{n} \left[ \prod_{s=\tau+1}^{n} (1-\gamma_{k,s}) \right]\gamma_{k,\tau} \left(F_k(\bm S_{\tau-1})-S_k^*\right)}_{D_{k,n}}\notag\\
	&+ \underbrace{\sum_{\tau=1}^{n} \left[ \prod_{s=\tau+1}^{n} (1-\gamma_{k,s}) \right] \gamma_{k,\tau}\bar{\xi}_{k,\tau}}_{E_{k,n}}.
\end{align}
We analyze the asymptotic behavior of these three terms:
\begin{itemize}
	\item \textbf{Term $C_{k,n}$ (Initial Condition Decay):} This term represents the decaying influence of the initial state. Since Lemma~\ref{lem:step_size} ensures $\sum_{\tau=1}^\infty \gamma_{k,\tau} = \infty$, the product term $\prod_{\tau=1}^\infty (1-\gamma_{k,\tau})$ vanishes, and thus $\lim_{n \to \infty} C_{k,n} = 0$ almost surely.
	
	\item \textbf{Term $E_{k,n}$ (Noise Attenuation):} This is a weighted sum of the martingale difference noise terms. Standard results in SA theory (e.g., Lemma 2 in \cite{tsitsiklis1994asynchronous}) show that the diminishing step sizes $\gamma_{k,\tau}$ are sufficient to average out the noise, yielding $\lim_{n \to \infty} E_{k,n} = 0$ almost surely.
	
	\item \textbf{Term $D_{k,n}$ (Mean Field Bias):} This term captures the persistent bias from the mean field dynamics. We can establish a uniform bound on this term by bounding the deviation $|F_k(\bm S) - S_k^*|$. For the posterior mean components (appeared in $F_k(\bm S)$),  this leads to the bounds:
	$|D_{i,n}| \le \max_{j \ne 1} |A_{i,j}-A_{i,1}|$ for Player 1's actions $i \in \Iscr$, and a symmetric bound for Player 2. For the variance components, $D_{k,n}=0$ since $w_k^*=v_k^*=0$ and $F_k(\bm S) \ge 0$.
\end{itemize}
Combining these results, we find that for any $\epsilon > 0$, there exists a random but finite $N_0$ such that for all $n \ge N_0$, the error is dominated by the bound on the mean field bias. For Player 1's posterior means, this implies:
$|x_{i,n} - A_{i,1}| \le \max_{j \ne 1}|A_{i,j} - A_{i,1}| + \epsilon$.
Now, the crucial role of Assumption~\ref{asp:payoff-matrix-mul} becomes clear. It is constructed precisely to ensure that this confinement for the iterates does not intersect the singular regions where $x_{i,n}\approx x_{k,n}$. By applying the triangle inequality and Assumption~\ref{asp:payoff-matrix-mul}, we can show that for $n \ge N_0$ and any distinct actions $i,k$:
\begin{align}
	|x_{i,n}-x_{k,n}| > \delta_1 > 0 \quad \text{and} \quad |y_{j,n}-y_{\ell,n}| > \delta_2 > 0,
\end{align}
for some positive constants $\delta_1, \delta_2$. This proves that the iterates are eventually bounded away from the regions where $F$ is not Lipschitz continuous.

\paragraph{Step 2: Establishing a Local Contraction Property.}
With the assurance that for $n > N_0$, the iterates $\bm S_n$ lie in a ``well-behaved'' set, we now show that the map $F$ acts as a contraction with respect to $\bm S^*$ in this region under the $\ell_{\infty}$ norm. We analyze the Jacobian of $F$ by applying the Mean Value Theorem:
\begin{align}
	F(\bm S_n) - F(\bm S^*) = \nabla F(\tilde{\bm S}_n) (\bm S_n -\bm S^*),
\end{align}
where $\tilde{\bm S}_n \coloneqq \big(\tilde{x}_{1,n},\dots, \tilde{x}_{I,n}, \tilde{y}_{1,n},\dots,\tilde{y}_{J,n}, \tilde{w}_{1,n},\dots,\tilde{w}_{I,n},\tilde{v}_{1,n},\dots,\tilde{v}_{J,n}\big)$ is a point on the line segment between $\bm S_n$ and $\bm S^*$. The goal is to show that the norm of the Jacobian matrix $\nabla F(\tilde{\bm S}_n)$ is less than 1 for large $n$.

The entries of the Jacobian involve derivatives of the choice probabilities (e.g., $\partial \varphi_i / \partial y_j$). These derivatives involve terms of the form $\exp(-\frac{1}{2} d^2)$, where $d$ is a z-score like $(\tilde{y}_{j,n}-\tilde{y}_{k,n})/\sqrt{\tilde{v}_{j,n}+\tilde{v}_{k,n}}$.
From Step 1, the numerator is bounded away from zero for $n > N_0$. From Lemma~\ref{lemma:infinite}, we know that all actions are played infinitely often, so the posterior variances in the denominator, $\tilde{v}_{j,n}$ and $\tilde{v}_{k,n}$, converge to zero. Consequently, the magnitude of the z-score $|d|$ tends to infinity. This drives the exponential term, and thus the off-diagonal elements of the Jacobian, to zero. A careful analysis shows that there exists some $\beta \in (0,1)$ such that for all $n$ sufficiently large we have the following contraction with respect to the $\ell_{\infty}$ norm:
\begin{align}
	\| F(\bm S_n) - F(\bm S^*)\|_\infty \leq \beta \| \bm S_n - \bm S^* \|_\infty.
\end{align}

\paragraph{Step 3: Concluding Convergence.}
Steps 1 and 2 provide all the necessary ingredients. We have a stochastic process $\{\bm S_n\}$ that, after a finite time $\eta_0$, evolves within a region where its mean field map $F$ is a contraction towards $\bm S^*$. This is a key condition that allows us to apply convergence results from the theory of asynchronous stochastic approximation, such as Theorem 3 in \cite{tsitsiklis1994asynchronous}. These results formally guarantee that the error term $\|\bm S_n - \bm S^*\|$ converges to zero almost surely.

\section{Numerical Studies} \label{sec:experiments}
In this section, we present results from simulation studies. We first present a numerical experiment to validate the theoretical results of Theorem \ref{thm:main-result}. We construct a 6×5 two-player game with expected payoff matrices $A_1$ and $B_1$ that satisfy Assumptions \ref{asp:unique-ne} and \ref{asp:payoff-matrix-mul}, which has a unique pure Nash equilibrium (action 1, action 1).
\begin{align}
	A_1 = \left( \begin{array}{ccccc}
		2.5 &1.9 &2.0 &1.9 &1.9\\
		0.1 &0.2 &0.3 &0.3 &0.2\\
		1.6 &1.8 &1.7 &1.8 &1.8\\
		0.5 &0.5 &0.4 &0.4 &0.4\\
		0.9 &0.8 &0.9 &0.8 &0.8\\
		1.2 &1.1 &1.2 &1.2 &1.1
	\end{array} \right),  \quad
	B_1 = \left( \begin{array}{ccccc}
		2.0 &0.5 &0.1 &1.0 &1.4\\
		1.8 &0.3 &0.2 &1.2 &1.3\\
		2.2 &0.4 &0.1 &1.1 &1.3\\
		1.9 &0.5 &0.1 &1.1 &1.4\\
		1.8 &0.3 &0.2 &1.2 &1.3\\
		1.9 &0.4 &0.2 &1.2 &1.4
	\end{array} \right).
\end{align}
The simulation employs Algorithm \ref{alg:ts1} over $10^6$ rounds. The players' prior beliefs for the expected payoff of each action are modeled as unit-variance normal distributions. The initial means for Player 1 are set to $\bm x_{0}=[0.8147, 0.9058, 0.1270, 0.9134, 0.6324, 0.0975]$ and for Player 2 to $\bm y_{0}=[0.2785, 0.5469, 0.9575, 0.9649, 0.1576]$. In each round, the realized payoff for a selected action is its true expected value plus additive noise drawn from a standard normal distribution. Figure \ref{fig:A1B1} depicts the probability of each player choosing their first action for a single, representative sample path. The horizontal axis is on a logarithmic scale to better visualize the convergence behavior. As shown, the probability of action 1 for both players converge to 1, which corresponds to the game's unique pure-strategy Nash equilibrium. This convergence is consistently observed across many simulations with randomly generated initial prior means, demonstrating the robustness of the outcome. These results provide strong numerical evidence for Theorem \ref{thm:main-result}, showing that even with misspecified initial beliefs, players' strategies converge to the Nash equilibrium under the specified assumptions.
\begin{figure}[tbp]
	\centering
	\begin{tikzpicture}
		\begin{axis}[
			xlabel={Time},
			ylabel={Probability},
			xlabel style={font=\small},
			ylabel style={font=\small},
			xmin=1, xmax=1000000,
			ymin=0, ymax=1.1,
			xmode=log,          
			log basis x=10,
			xtick={1,100,10000,1000000},
			xticklabels={1,$10^2$,$10^4$,$10^6$},
			ytick={0,0.2,0.4,0.6,0.8,1},
			xticklabel style={font=\small},
			yticklabel style={font=\small},
			legend pos=south east,
			legend style={font=\small},
			grid=minor
			]
			\addplot+[solid, thick, mark=none] table [x=Time, y=Phi, col sep=comma] {Data/TS1-1.csv};
			\addlegendentry{Player 1 Action 1}
			\addplot+[dashed, thick, mark=none] table [x=Time, y=Psi, col sep=comma] {Data/TS1-1.csv};
			\addlegendentry{Player 2 Action 1}
		\end{axis}
	\end{tikzpicture}
	\caption{The probability that the two players choose action 1 with payoff matrices $(A_1,B_1)$ for a representative sample path}
	\label{fig:A1B1}
\end{figure}

We next consider a game that satisfies Assumption~\ref{asp:unique-ne} but not Assumption~\ref{asp:payoff-matrix-mul}. 
The example is the Prisoner's Dilemma presented earlier in Section~\ref{sec:motivating-exp}.
We set prior distributions for each player's actions using randomly generated means. 
The prior distributions for Player 1 are $\mathcal{N}(4.0736, 1)$ for action 1 and $\mathcal{N}(4.5290, 1)$ for action 2. 
For Player 2, the priors are $\mathcal{N}(0.6349, 1)$ and $\mathcal{N}(4.5669, 1)$ for actions 1 and 2, respectively. 
The experiment consists of 500 independent runs, each lasting $10^6$ rounds. 
As we have seen in Figure~\ref{fig:ts-collude}, 
22.6\% of the simulated paths converge to the collusive outcome instead of the Nash equilibrium.

Next, we investigate the robustness of our approach to model misspecification, where the true reward-generating distribution deviates from the distribution assumed by the learning algorithm. We design a $2 \times 2$ game with payoff matrices $A_2$ and $B_2$ whose unique pure-strategy Nash equilibrium is the action profile $(1,1)$.
\begin{align}
	A_2 = \left( \begin{array}{cc}
		0.5 &0.4\\
		0.2 &0.3
	\end{array} \right), \quad
	B_2 = \left( \begin{array}{cc}
		0.7 &0.3\\
		0.6 &0.5
	\end{array} \right).
\end{align}
In this experiment, the realized rewards are drawn from a Bernoulli distribution, with the success probabilities given by the entries of the expected payoff matrices. However, the players implement Algorithm \ref{alg:ts1}, which incorrectly assumes a unit-variance normal reward distribution for its Bayesian updates. Player 1 has prior distributions $\mathcal{N}(0.8147,1)$ for action 1 and $\mathcal{N}(0.9058,1)$ for action 2, while player 2 has prior distributions $\mathcal{N}(0.1270,1)$ for action 1 and $\mathcal{N}(0.9134,1)$ for action 2. Despite this significant model misspecification, the system converges to the Nash Equilibrium. Figure \ref{fig:A2B2} illustrates the probability of each player selecting action 1 for a representative sample path. The probability for both players converge to 1, consistent with the NE. This result is consistently observed across multiple simulations, suggesting that the convergence guarantees of Thompson sampling in our game are robust even when the distributional settings are violated.
\begin{figure}[tbp]
	\centering
	\begin{tikzpicture}
		\begin{axis}[
			xlabel={Time},
			ylabel={Probability},
			xlabel style={font=\small},
			ylabel style={font=\small},
			xmin=1, xmax=1000000,
			ymin=0, ymax=1.1,
			xmode=log,          
			log basis x=10,
			xtick={1,100,10000,1000000},
			xticklabels={1,$10^2$,$10^4$,$10^6$},
			ytick={0,0.2,0.4,0.6,0.8,1},
			xticklabel style={font=\small},
			yticklabel style={font=\small},
			legend pos=south east,
			legend style={font=\small},
			grid=minor
			]
			\addplot[solid, thick, blue] table [x=Time, y=Phi, col sep=comma] {Data/TS3-1.csv};
			\addlegendentry{Player 1 Action 1}
			\addplot[dashed, thick, red] table [x=Time, y=Psi, col sep=comma] {Data/TS3-1.csv};
			\addlegendentry{Player 2 Action 1}
		\end{axis}
	\end{tikzpicture}
	\caption{The probability that the two players choose action 1 with payoff matrices $(A_2,B_2)$ for a representative sample path}
	\label{fig:A2B2}
\end{figure}

Finally, we conduct two experiments to demonstrate the system's dynamics when Assumption \ref{asp:unique-ne} is violated, meaning the game does not have a unique pure-strategy Nash equilibrium (NE). For these simulations, the prior belief for each action's expected payoff is a standard normal distribution, $\mathcal{N}(0,1)$. We consider two cases given below:
\begin{align}
	A_3 = \left( \begin{array}{cc}
		0.3 &0.3\\
		0.4 &0.1
	\end{array} \right), \quad
	B_3 = \left( \begin{array}{cc}
		0.1 &0.3\\
		0.4 &0.3
	\end{array} \right); \quad
	A_4 = \left( \begin{array}{cc}
		0.5 &0.2\\
		0.1 &0.3
	\end{array} \right), \quad
	B_4 = \left( \begin{array}{cc}
		0.3 &0.5\\
		0.7 &0.4
	\end{array} \right).
\end{align}
\begin{itemize}
	\item[(i)] Two pure-strategy NE and one mixed-strategy NE. We first consider a 2×2 game with payoff matrices $(A_3, B_3)$ that possesses two pure-strategy NE (action 1, action 2), (action 2, action 1) and one mixed-strategy NE $(1/3, 2/3)$, which means that two players choose action 1 with probability $1/3$ and $2/3$ respectively. We simulate 2,000 independent sample paths to observe the long-term outcomes. We find that the dynamics consistently converge to one of the two pure-strategy equilibria, effectively selecting one of them based on the stochastic learning path. Across the simulations, the system converged to the (action 1, action 2) equilibrium in 80\% of the runs and to the (action 2, action 1) equilibrium in the remaining 20\%. This demonstrates that when multiple pure-strategy NE exist, the learning process does not oscillate but eventually settles on one of them.
	\item[(ii)] No pure-strategy NE and one mixed-strategy NE. In the second case, we use payoff matrices $(A_4,B_4)$ for a game that has no pure-strategy NE and a mixed-strategy NE $(0.6, 0.2)$, which means that two players choose action 1 with probability 0.6 and 0.2, respectively.
	The results, illustrated in Figure \ref{fig:mixed}, reveal a different dynamic behavior. 
	We observe that for each player, the posterior means of the payoffs for their two actions converge to nearly identical values. This is consistent with the condition for a mixed strategy, where a player must be indifferent between the actions they play with positive probability. However, the probability of choosing action 1 do not converge to a stable value. Instead, they exhibit persistent oscillations around the mixed-strategy probabilities. This oscillating behavior arises because when a player's posterior means are nearly equal, the Thompson sampling choice becomes highly sensitive to small updates. A single noisy reward can be enough to shift the posterior belief slightly, causing the probability of selecting that action to swing from near zero to near one. 
	This instability prevents the players' strategies from stabilizing, leading to the observed cyclical dynamics.
\end{itemize}
\begin{figure}[tbp]
	\centering
	\begin{subfigure}[t]{8cm}
		\begin{tikzpicture}[scale=0.9]
			\begin{axis}[
				xlabel={Time},
				ylabel={Posterior Mean},
				xlabel style={font=\small},
				ylabel style={font=\small},
				xmin=1000000, xmax=10000000,
				ymin=0, ymax=1.3,
				xmode=log,          
				log basis x=10,
				xtick={1000000,5000000,10000000},
				xticklabels={$10^6$,$5\times10^6$,$10^7$},
				ytick={0,0.2,0.4,0.6,0.8,1},
				xticklabel style={font=\small},
				yticklabel style={font=\small},
				legend pos=north east,
				legend style={font=\small},
				grid=minor
				]
				\addplot[solid, thick, blue] table [x=Time, y=x1, col sep=comma] {Data/TS4-2.csv};
				\addlegendentry{Player 1 Action 1}
				\addplot[dashed, thick, OliveGreen] table [x=Time, y=x2, col sep=comma] {Data/TS4-2.csv};
				\addlegendentry{Player 1 Action 2}
				\addplot[solid, thick, YellowOrange] table [x=Time, y=y1, col sep=comma] {Data/TS4-2.csv};
				\addlegendentry{Player 2 Action 1}
				\addplot[dashed, thick, red] table [x=Time, y=y2, col sep=comma] {Data/TS4-2.csv};
				\addlegendentry{Player 2 Action 2}
			\end{axis}
		\end{tikzpicture}
		\caption{\footnotesize{The posterior means of each action}}
	\end{subfigure}
	\begin{subfigure}[t]{8cm}
		\centering
		\begin{tikzpicture}[scale=0.9]
			\begin{axis}[
				xlabel={Time},
				ylabel={Probability},
				xlabel style={font=\small},
				ylabel style={font=\small},
				xmin=1000000, xmax=10000000,
				ymin=0, ymax=1.3,
				xmode=log, 
				log basis x=10,
				xtick={1000000,5000000,10000000},
				xticklabels={$10^6$,$5 \times 10^6$, $10^7$},
				ytick={0,0.2,0.4,0.6,0.8,1},
				xticklabel style={font=\small},
				yticklabel style={font=\small},
				legend pos=north east,
				legend style={font=\small},
				grid=minor
				]
				\addplot[solid, thick, blue] table [x=Time, y=Phi, col sep=comma] {Data/TS4-2.csv};
				\addlegendentry{Player 1 Action 1}
				\addplot[dashed, thick, red] table [x=Time, y=Psi, col sep=comma] {Data/TS4-2.csv};
				\addlegendentry{Player 2 Action 1}
			\end{axis}
		\end{tikzpicture}
		\caption{\footnotesize{The probability of choosing action 1}}
	\end{subfigure}
	\caption{A game with no pure-strategy NE and one mixed-strategy NE, given the payoff matrices $A_4$ and $B_4$.}
	\label{fig:mixed}
\end{figure}

\section{Conclusion}\label{sec:conclusion}
In this paper, we study repeated games where agents independently employ Thompson sampling to navigate the trade-off between exploration and exploitation. Our central goal is to address whether this common, single-agent learning algorithm could lead to unintentional algorithmic collusion when deployed in a multi-agent setting. We establish mild conditions on the game's payoff matrices that guarantee the players' strategies converge to the unique pure-strategy Nash Equilibrium. This finding provides a definitive theoretical resolution to the puzzle of emergent collusion, demonstrating that, under these conditions, competitive outcomes prevail. 

A central contribution of this work is methodological. Proving our main convergence result requires developing a novel, sample-path-wise proof technique, as the system's dynamics defy analysis by standard stochastic approximation frameworks. We also validate and explore the boundaries of this theory through extensive numerical experiments. These simulations not only verify convergence under the proposed assumptions but also demonstrate that violating them can lead to collusive outcomes or non-convergent oscillations, highlighting the critical role of the game's payoff structure.

\bibliographystyle{plainnat}
\bibliography{ref} 

\newpage
\appendix
\section{Probability of Action Selection}\label{sec:prob}
To obtain the closed form of the probability $\varphi_{i,n+1}$, recall that
\begin{align}\label{equ:phi-mul-2}
	\varphi_{i,n+1}=\PR(i_{n+1}=i|\mathcal{F}_n)= \PR(\theta_{i,n+1}>\max_{k \ne i}\theta_{k,n+1}|\bm S_n).
\end{align}
Let us consider a random vector
\begin{align}
	(\theta_{i,n+1}-\theta_{1,n+1}, \dots, \theta_{i,n+1}-\theta_{i-1,n+1}, \theta_{i,n+1}-\theta_{i+1,n+1}, \dots, \theta_{i,n+1}-\theta_{I,n+1}) \in \mathbb{R}^{I-1},
\end{align}
which follows a multivariate normal distribution with mean $(x_{i,n}-x_{1,n},\dots, x_{i,n}-x_{I,n})$ and covariance matrix
\begin{align}
	\left( \begin{array}{cccc}
		w_{i,n}+w_{1,n} &w_{i,n} &\cdots &w_{i,n}\\
		w_{i,n} &w_{i,n}+w_{2,n} &\cdots &w_{i,n}\\
		\vdots &\vdots &\ddots  &\vdots\\
		w_{i,n} &w_{i,n} &\cdots & w_{i,n}+w_{I,n}
	\end{array} \right).
\end{align}
Then we can further calculate \eqref{equ:phi-mul-2} to
\begin{align}
	\varphi_{i,n+1}
	&= \PR\left (\theta_{i,n+1}-\theta_{1,n+1} \geq 0, \dots, \theta_{i,n+1}-\theta_{i-1,n+1} \geq 0, \theta_{i,n+1}-\theta_{i+1,n+1} \geq 0, \theta_{i,n+1}-\theta_{I,n+1} \geq 0|\bm S_n \right )\\
	&=\PR \left(Z_{1,n+1}^{(i)} \leq c_{1,n}^{(i)}, \dots, Z_{i-1,n+1}^{(i)} \leq c_{i-1,n}^{(i)}, Z_{i+1,n+1}^{(i)} \leq c_{i+1,n}^{(i)}, \dots, Z_{I,n+1}^{(i)} \leq c_{I,n}^{(i)} \Big|\bm S_n \right)\\
	&=\int_{-\infty}^{c_{1,n}^{(i)}}\dots \int_{-\infty}^{c_{I,n}^{(i)}} \phi(z_1, \cdots,z_I) dz_I\cdots dz_1\\
	&= \Phi_{I-1}\left((c_{k,n}^{(i)})_{k \ne i}; \bm \rho_{n}^{(i)}\right),\label{equ:phi-mul-3}
\end{align}
where $c_{k,n}^{(i)}=\frac{x_{i,n}-x_{k,n}}{\sqrt{w_{i,n}+w_{k,n}}}$ for $k \ne i$. In the second equation, $(Z_{1,n+1}^{(i)} ,\dots, Z_{i-1,n+1}^{(i)},Z_{i+1,n+1}^{(i)},\dots,Z_{I,n+1}^{(i)} )$ follows a multivariate normal distribution $N(\bm 0, \bm \rho_{n}^{(i)})$ with correlation matrix
\begin{align}
	\bm \rho_{n}^{(i)}=\left( \begin{array}{cccc}
		1 &\frac{w_{i,n}}{\sqrt{(w_{i,n}+w_{1,n})(w_{i,n}+w_{2,n})}} &\cdots &\frac{w_{i,n}}{\sqrt{(w_{i,n}+w_{1,n})(w_{i,n}+w_{I,n})}}\\
		\frac{w_{i,n}}{\sqrt{(w_{i,n}+w_{2,n})(w_{i,n}+w_{1,n})}} &1 &\cdots &\frac{w_{i,n}}{\sqrt{(w_{i,n}+w_{2,n})(w_{i,n}+w_{I,n})}}\\
		\vdots &\vdots &\ddots  &\vdots\\
		\frac{w_{i,n}}{\sqrt{(w_{i,n}+w_{I,n})(w_{i,n}+w_{1,n})}} &\frac{w_{i,n}}{\sqrt{(w_{i,n}+w_{I,n})(w_{i,n}+w_{2,n})}} &\cdots & 1
	\end{array} \right) \in \mathbb{R}_{+}^{(I-1) \times (I-1)},
\end{align}
and the probability density function $\phi(z_1, \cdots, z_{i-1}, z_{i+1}, z_I)$. In the last equation, $\Phi_{I-1}$ is the CDF of a $(I-1)$-variate standard normal distribution with correlation matrix $\bm \rho_{n}^{(i)}$.

Similarly, we can also calculate the probability $\psi_{j,n+1}$ by 
\begin{align}\label{equ:psi-mul}
	\psi_{j,n+1}\coloneqq \psi_{j}(\bm S_n) \coloneqq\PR( j_{n+1}=j|\mathcal{F}_n) =\PR(\vartheta_{j,n+1} \geq \max_{k \ne j} \vartheta_{k,n+1} | \bm S_n) = \Phi_{J-1} \left ( \left(d_{k,n}^{(j)} \right)_{ k \ne j};  \bm \varrho_{j,n}\right),
\end{align}
where $d_{k,n}^{(j)} \coloneqq  \frac{y_{j,n}-y_{k,n}}{\sqrt{v_{j,n}+v_{k,n}}}$ for $k \ne j$ and 
\begin{align}
	\bm\varrho_{j,n}=\left( \begin{array}{cccc}
		1 &\frac{v_{j,n}}{\sqrt{(v_{j,n}+v_{1,n})(v_{j,n}+v_{2,n})}} &\cdots &\frac{v_{j,n}}{\sqrt{(v_{j,n}+v_{1,n})(v_{j,n}+v_{J,n})}}\\
		\frac{v_{j,n}}{\sqrt{(v_{j,n}+v_{2,n})(v_{j,n}+v_{1,n})}} &1 &\cdots &\frac{v_{j,n}}{\sqrt{(v_{j,n}+v_{2,n})(v_{j,n}+v_{J,n})}}\\
		\vdots &\vdots &\ddots  &\vdots\\
		\frac{v_{j,n}}{\sqrt{(v_{j,n}+v_{J,n})(v_{j,n}+v_{1,n})}} &\frac{v_{j,n}}{\sqrt{(v_{j,n}+v_{J,n})(v_{j,n}+v_{2,n})}} &\cdots & 1
	\end{array} \right) \in \mathbb{R}_{+}^{(J-1) \times (J-1)}.
\end{align}


\section{Proof of Lemma~\ref{lemma:infinite}}\label{sec:proof-lemma-infinite}
\begin{proof}{Proof of Lemma \ref{lemma:infinite}}
	The lemma states that for any player, each of their actions is chosen infinitely often almost surely. We provide the proof for Player 1 and action $i=1$, i.e., we show $\lim_{n \to \infty} N_{1,n}=\infty$ almost surely. The proofs for any other action $k \neq 1$ and for Player 2 are identical.
	
	Our proof strategy is by contradiction, using the second Borel-Cantelli Lemma. Let $E_{1,n}$ be the event that Player 1 chooses action 1 in round $n$, so $E_{1,n} = \{i_n = 1\}$. The event that action 1 is chosen infinitely often is precisely ${E_{1,n} \text{ i.o.}}$. We have the equivalence:
	\begin{align}\label{N_equality1}
		\left\{\lim_{n \to \infty} N_{1,n}=\infty \right\} = \left\{E_{1,n} \text{ occurs infinitely often}\right\}.
	\end{align}
	The event $E_{1,n}$ is measurable with respect to the filtration $\mathcal{F}_n$. The conditional version of the second Borel-Cantelli Lemma (see, e.g., Theorem 5.3.2 in \cite{durrett2019probability}) states that
	\begin{align}\label{N_equality2}
		\left\{\sum_{n=1}^{\infty} \PR(E_{1,n}|\mathcal{F}_{n-1})=\infty \right\} \subseteq \left\{E_{1,n} \text{ i.o.}\right\} \quad \text{a.s.}
	\end{align}
	We define $\varphi_{1,n}\coloneqq \PR(E_{1,n}|\mathcal{F}_{n-1})  $. Therefore, to prove our claim, it is sufficient to show that $\sum_{n=1}^{\infty} \varphi_{1,n} = \infty$ almost surely.
	
	We proceed by contradiction. Assume the contrary, that the set of sample paths where action 1 is chosen only a finite number of times has positive probability. Let $\Omega_{\text{finite}}$ be this set of paths:
	\begin{align}
		\Omega_{\text{finite}} \coloneqq \left\{ \bm{\omega} : \lim_{n \to \infty} N_{1,n}(\bm{\omega}) < \infty \right\}.
	\end{align}
	Our assumption is $\PR(\Omega_{\text{finite}}) > 0$. Let us analyze the behavior of $\sum \varphi_{1,n}$ on an arbitrary sample path $\bm{\omega} \in \Omega_{\text{finite}}$. On such a path, there exists a finite integer $\overline{N}_1(\bm{\omega})$ such that $N_{1,n}(\bm{\omega}) \le \overline{N}_1(\bm{\omega})$ for all $n$.
	Our goal is to show that for any such path $\bm{\omega}$, the sum $\sum_{n=1}^{\infty} \varphi_{1,n}(\bm{\omega})$ diverges. This will lead to the desired contradiction.
	
	From the analysis in Appendix~\ref{sec:prob}, we can lower bound $\varphi_{1,n}$ using Slepian's inequality, as the off-diagonal entries of the correlation matrix $\rho_n^{(1)}$ are non-negative:
	\begin{align}\label{inequ:phi-mul-lower}
		\varphi_{1,n} \geq \prod_{k=2}^{I} \Phi\left( \frac{x_{1,n-1}-x_{k,n-1}}{\sqrt{w_{1,n-1}+w_{k,n-1}}} \right).
	\end{align}
	We will show that for our path $\bm{\omega}$, each term in this product is bounded below by a positive constant that is independent of $n$. For any action $k \in \{2, \dots, I\}$, we consider two cases at each step $n-1$:
	
	\textbf{Case 1: $x_{1,n-1}(\bm{\omega}) \geq x_{k,n-1}(\bm{\omega})$.}
	In this case, the argument of the CDF $\Phi(\cdot)$ is non-negative, so
	\begin{align}\label{inequ:Phi-mul-case1}
		\Phi\left( \frac{x_{1,n-1}-x_{k,n-1}}{\sqrt{w_{1,n-1}+w_{k,n-1}}} \right) \geq \Phi(0) = \frac{1}{2}.
	\end{align}
	
	\textbf{Case 2: $x_{1,n-1}(\bm{\omega}) < x_{k,n-1}(\bm{\omega})$.}
	Let $\lambda_{k,n-1} \coloneqq \frac{x_{k,n-1}-x_{1,n-1}}{\sqrt{w_{1,n-1}+w_{k,n-1}}} > 0$. The term becomes $\Phi(-\lambda_{k,n-1}) = 1 - \Phi(\lambda_{k,n-1})$. Using the standard lower bound for the normal CDF tail (e.g., Formula 7.1.13 in \cite{abramowitz1948handbook}), for any $x \ge 0$, $1-\Phi(x) \ge \frac{1}{\sqrt{2\pi}} \frac{1}{x+\sqrt{x^2+4}} e^{-x^2/2}$. This gives:
	\begin{align}\label{inequ:Phi-mul-case2}
		\Phi(-\lambda_{k,n-1}) \geq \frac{1}{\sqrt{2\pi}} \frac{1}{\lambda_{k,n-1}+\sqrt{\lambda_{k,n-1}^2+4}} e^{-\lambda_{k,n-1}^2/2}.
	\end{align}
	To get a uniform lower bound, we need to show that $\lambda_{k,n-1}(\bm{\omega})$ is bounded above for all $n$ on the path $\bm{\omega}$.
	
	\textit{Bounding the numerator $x_{k,n-1}-x_{1,n-1}$:}
	The reward at step $s$ is $a_{i_s,s} = A_{i_s,j_s} + \xi_s$, where ${\xi_s}$ are i.i.d. $\mathcal{N}(0,1)$. Let $\bar{A}=\max_{i,j} |A_{i,j}|$.
	\begin{align*}
		|x_{k,n-1}| &= \left|\frac{x_{k,0}+\sum_{s=1}^{n-1} a_{i_s,s} \cdot \I{i_s=k}}{N_{k,n-1}+1}\right| \le \frac{|x_{k,0}| + N_{k,n-1}\bar{A} + \sum_{s=1}^{n-1} |\xi_s| \cdot \I{i_s=k}}{N_{k,n-1}+1}.
	\end{align*}
	On the path $\bm{\omega}$, for any action $k$, either $N_{k,n}(\bm{\omega}) \to \infty$ or $N_{k,n}(\bm{\omega})$ converges to a finite limit.
	\begin{itemize}
		\item If $N_{k,n}(\bm{\omega}) \to \infty$, then by the Strong Law of Large Numbers, $\frac{\sum_{s=1}^{n-1} |\xi_s| \cdot \I{i_s=k}}{N_{k,n-1}} \to \mathbb{E}[|\xi_1|] = \sqrt{2/\pi}$. Thus, the sequence ${|x_{k,n-1}(\bm{\omega})|}_{n \ge 1}$ is bounded.
		\item If $N_{k,n}(\bm{\omega})$ is finite, the numerator is a sum of a finite number of terms, so the sequence ${|x_{k,n-1}(\bm{\omega})|}_{n \ge 1}$ is also bounded.
	\end{itemize}
	Since this holds for all $k$, including $k=1$, the difference $x_{k,n-1}(\bm{\omega})-x_{1,n-1}(\bm{\omega})$ is bounded in $n$ by some constant $C_k(\bm{\omega})$.
	
	\textit{Bounding the denominator $\sqrt{w_{1,n-1}+w_{k,n-1}}$:}
	On the path $\bm{\omega} \in \Omega_{\text{finite}}$, we have $N_{1,n-1}(\bm{\omega}) \le \overline{N}_1(\bm{\omega})$. Therefore, $w_{1,n-1}(\bm{\omega}) = \frac{1}{N_{1,n-1}(\bm{\omega})+1} \ge \frac{1}{\overline{N}_1(\bm{\omega})+1} > 0$. Since $w_{k,n-1} > 0$, the denominator is bounded below:
	\begin{align*}
		\sqrt{w_{1,n-1}(\bm{\omega})+w_{k,n-1}(\bm{\omega})} \ge \sqrt{w_{1,n-1}(\bm{\omega})} \ge \frac{1}{\sqrt{\overline{N}_1(\bm{\omega})+1}} > 0.
	\end{align*}
	Combining these, $\lambda_{k,n-1}(\bm{\omega})$ is bounded above for all $n$ by some constant $\bar{C}_k(\bm{\omega}) < \infty$. Substituting this into \eqref{inequ:Phi-mul-case2} gives a positive lower bound for Case 2 that depends on $\bm{\omega}$ but not on $n$.
	
	\textit{Conclusion for the path $\bm{\omega}$:}
	For any $k \neq 1$ and any $n$, either \eqref{inequ:Phi-mul-case1} or \eqref{inequ:Phi-mul-case2} holds. We can define a path-dependent constant $\bar{\varphi}_k(\bm{\omega}) > 0$ as
	\begin{align}
		\bar{\varphi}_k(\bm{\omega}) \coloneqq \min\left\{\frac{1}{2}, \frac{1}{\sqrt{2\pi}} \frac{e^{-\bar{C}k(\bm{\omega})^2/2}}{\bar{C}k(\bm{\omega})+\sqrt{\bar{C}k(\bm{\omega})^2+4}}\right\}.
	\end{align}
	This ensures that for all $n$, $\Phi\left( \frac{x_{1,n-1}-x_{k,n-1}}{\sqrt{w_{1,n-1}+w_{k,n-1}}} \right) \ge \bar{\varphi}_k(\bm{\omega})$.
	Plugging this back into \eqref{inequ:phi-mul-lower}, we find that for all $n$ on the path $\bm{\omega}$,
	\begin{align}
		\varphi_{1,n}(\bm{\omega}) \geq \prod_{k=2}^{I} \bar{\varphi}_k(\bm{\omega}) > 0.
	\end{align}
	Letting the product be $\bar{\varphi}(\bm{\omega})^{I-1}$, we have $\sum_{n=1}^{\infty} \varphi_{1,n}(\bm{\omega}) \geq \sum_{n=1}^{\infty} \bar{\varphi}(\bm{\omega})^{I-1} = \infty$.
	
	We have shown that for any sample path $\bm{\omega} \in \Omega_{\text{finite}}$, the sum $\sum_{n=1}^{\infty} \varphi_{1,n}(\bm{\omega})$ diverges. This implies:
	\begin{align}
		\Omega_{\text{finite}} = \left\{\lim_{n \to \infty} N_{1,n} < \infty\right\} \subseteq \left\{\sum_{n=1}^{\infty} \varphi_{1,n}=\infty\right\}.
	\end{align}
	However, by the second Borel-Cantelli Lemma, the set on the right is, up to a set of measure zero, a subset of ${\lim_{n \to \infty} N_{1,n} = \infty}$.
	\begin{align}
		\left\{\sum_{n=1}^{\infty} \varphi_{1,n}=\infty\right\} \subseteq \left\{\lim_{n \to \infty} N_{1,n} = \infty\right\} \quad \text{a.s.}
	\end{align}
	This means $\Omega_{\text{finite}}$ is a subset of its own complement, which is only possible if $\PR(\Omega_{\text{finite}})=0$. This contradicts our initial assumption that $\PR(\Omega_{\text{finite}}) > 0$.

	Therefore, the event ${\lim_{n \to \infty} N_{1,n}=\infty}$ must hold with probability one.
\end{proof}
\section{Proof of Lemma~\ref{lem:step_size}}\label{sec:proof-lemma-stepsize}
\begin{proof}{Proof of Lemma~\ref{lem:step_size}.}
	We show that for $i \in \mathcal{I}$, $\sum_{n=1}^{+\infty} \alpha_{i,n}=\infty$ and $\sum_{n=1}^{+\infty} \alpha_{i,n}^2<\infty$ almost surely. The arguments for $(\beta_{j,n})_{j \in \mathcal{J}}$ are completely analogous. Recall from Section~\ref{sec:state-dynamics} that $\alpha_{i,n}$ are binary-valued random variables with $\alpha_{i,n}=\frac{1}{N_{i,n}+1}$ if action $i$ is selected by Player 1 in round $n$ and $\alpha_{i,n}=0$ otherwise. Fix any sample path in the probability one set where Lemma~\ref{lemma:infinite} holds. 
	
	Suppose at time $s$, action $i$ is chosen and $N_{i,s-1}=a$ for some $a \in \mathbb{N}$, then we have $N_{i,s}=a+1$ and $\alpha_{i,s}=\frac{1}{N_{i,s}+1}=\frac{1}{a+2}$ on this path. Lemma~\ref{lemma:infinite} shows that action $i$ is chosen infinitely often, so there exists some time $\tau>s$ that action $i$ is chosen again, and  $\alpha_{i,\tau}=\frac{1}{N_{i,\tau}+1}=\frac{1}{(a+2)+1}=\frac{1}{a+3}$. 
	Repeating this argument, we can infer that $\left\{\frac{1}{n}\right\}_{n=1}^{+\infty}$ is a subsequence of $\{\alpha_{i,n}\}_{n=1}^\infty$ on such a sample path, where other elements in the sequence $\{\alpha_{i,n}\}_{n=1}^\infty$ are all zero. Therefore, we obtain $\sum_{n=1}^{+\infty} \alpha_{i,n}= \sum_{n=1}^{+\infty} \frac{1}{n}=\infty$ and $\sum_{n=1}^{+\infty} \alpha_{i,n}^2 = \sum_{n=1}^{+\infty} \frac{1}{n^2}<\infty$ on such a sample path. This completes the proof of Lemma~\ref{lem:step_size}. 
\end{proof}


\section{Proof of Lemma~\ref{lem:martingale_diff}}\label{sec:proof-lemma-martingale_diff}
\begin{proof}{Proof of Lemma~\ref{lem:martingale_diff}.}
	From the derivation in Section~\ref{sec:state-dynamics}, it is straightforward to verify that it is a Martingale difference.
	We next verify that it is square-integrable.
	We can directly compute from \eqref{bar-a} that 
	\begin{align}
		\E[\bar{a}_{i, n}^2|\Fscr_{n-1}]&= 1 +  \E \left[ \left(\sum_{j=1}^J A_{i,j} \1_{\{j_{n}=j\}} - \sum_{j=1}^J A_{i,j} \psi_{j,n} \right)^2 \big|\Fscr_{n-1} \right]\\
		&=1+\sum_{j=1}^{J} A_{ij}^2 \psi_{j,n}-\left(\sum_{j=1}^{J} A_{ij}\psi_{j,n}\right)^2\\
		&=1+\sum_{j=1}^{J} A_{ij}^2 \psi_{j,n}(1-\psi_{j,n})-\sum_{j \ne j'} A_{ij} A_{ij'} \psi_{j,n} \psi_{j',n}\\
		&\leq 1+ \sum_{j=1}^{J} A_{ij}^2/4,
	\end{align}
	where the second equality follows from the fact that given $\Fscr_{n-1}$, $\1_{\{j_{n}=j\}}$ is a Bernoulli random variable, and that $\E[ 1_{\{j_{n}=j\}}1_{\{j_{n}=k\}} | \Fscr_{n-1}]=0 $ for $j \ne k$. 
	
	Similarly, we have
	\begin{align}
		\E[\bar{b}_{j, n}^2|\Fscr_{n-1}]& = 1 +  \E \left[ \left(\sum_{i=1}^I B_{i,j} \1_{\{i_{n}=i\}} - \sum_{i=1}^I B_{i,j} \varphi_{i,n}\right)^2 \big|\Fscr_{n-1} \right]
		\le 1 + \sum_{i=1}^{I} B_{ij}^2/4. 
	\end{align}
	Therefore we obtain from \eqref{eq:noise-mul} that for all $n,$
	\begin{align}
		E\left[\bar{\bm \xi}_{n}^2|\Fscr_{n-1}\right]
		&= \sum_{i=1}^I \E \left[\bar{a}_{i, n}^2|\Fscr_{n-1}\right] +  \sum_{j=1}^J \E \left[\bar{b}_{j, n}^2|\Fscr_{n-1}\right]
		\le I+J + \sum_{i=1}^I \sum_{j=1}^J (A_{i,j}^2+ B_{i,j}^2)/4. 
	\end{align}
	The proof is then complete. 
\end{proof}

\section{Proof of Lemma~\ref{lem:bounded}}\label{sec:proof-lemma-bounded}
\begin{proof}{Proof of Lemma~\ref{lem:bounded}.}
	Recall that $\bm S_n= \big(x_{1,n},\dots, x_{I,n}, y_{1,n},\dots,y_{J,n}, w_{1,n},\dots,w_{I,n},v_{1,n},\dots,v_{J,n}\big)\in \R^{I+J} \times \R_+^{I+J}$. From the definitions in \eqref{eq:x-w}, we obtain that and $|w_{i,n}|$ and $|v_{j,n}|$ are bounded above by 1 for $i \in \mathcal{I}, j \in \mathcal{J}$. Then we have 
	\begin{align}
		\sup_{n} \Vert \bm S_n \Vert \leq 	\sum_{i=1}^I \sup_{n} |x_{i,n}| + 	\sum_{j=1}^J\sup_{n} |y_{j,n}| + I+J.
	\end{align}
	
	We first prove that $\sup_{n} |x_{i,n}| < \infty$ almost surely for $i \in \mathcal{I}$. Recall from \eqref{eq:x-w} that 
	\begin{align} 
		x_{i,n}=\frac{x_{i,0}+\sum_{s=1}^n a_{i_s,s} \cdot \1_{\{i_s=i\}}}{N_{i,n}+1},
	\end{align}
	where $a_{i_s,s}$ is the random reward following a normal distribution $\mathcal{N}(A_{i_s,j_s},1)$, and $N_{i,n}=\sum_{s=1}^n \1_{\{i_s=i\}}$ denotes the number of plays of action $i$ by Player 1 up to round $n$. We can write $a_{i_s,s} = A_{i_s,j_s} + \xi_s$,
	where $(\xi_s: s \ge 1 )$ is a sequence of i.i.d. standard normal random variables.
	Hence, we have
	\begin{align} 
		&|x_{i,n}| \le\frac{|x_{i,0}|+\sum_{s=1}^n |a_{i_s,s} | \cdot \1_{\{i_s=i\}}}{N_{i,n}+1}\le |x_{i,0}|+\max_{i,j} |A_{i,j}| + \frac{\sum_{s=1}^n |\xi_s | \cdot \1_{\{i_s=i\}} }{N_{i,n}+1}. 
	\end{align}
	It follows that
	\begin{align} \label{eq:sup-xn}
		\sup_{n} |x_{i,n}| 
		\le |x_{i,0}|+\max_{i,j} |A_{i,j}| + \sup_{n} \frac{\sum_{s=1}^n |\xi_s | \cdot \1_{\{i_s=i\}} }{N_{i,n}+1}.
	\end{align}
	
	For each sample path, $\{N_{i,n}: n \ge 1\}$ is a non-decreasing sequence of integers and hence we can set $N_{i,\infty} = \lim_{n \rightarrow \infty} N_{i,n}$, which can be $\infty$. 
	In view of \eqref{eq:sup-xn}, to show $\sup_{n} |x_{i,n}| < \infty$ almost surely, it suffices to consider those sample paths with $N_{i,\infty} = \infty.$ For each of such sample paths (except a possible zero-probability set), it holds $\frac{\sum_{s=1}^n |\xi_s | \cdot \1_{\{i_s=i\}} }{N_{i,n}+1}<\frac{\sum_{s=1}^n |\xi_s | \cdot \1_{\{i_s=i\}} }{N_{i,n}}$. We can infer from the strong law of large numbers that $\lim_{n \rightarrow \infty} \frac{\sum_{s=1}^n |\xi_s | \cdot \1_{\{i_s=i\}} }{N_{i,n}} = \E[|\xi_1|] <\infty$. This implies that $\sup_{n} \frac{\sum_{s=1}^n |\xi_s | \cdot \1_{\{i_s=i\}} }{N_{i,n}+1} < \infty$ on such paths. Therefore, we can infer from \eqref{eq:sup-xn} that $ \sup_{n} |x_{i,n}| <\infty$ almost surely.
	
	Similarly, we can prove that $\sup_n |y_{j,n}| <\infty$ almost surely for $j\in \mathcal{J}$. Thus, we obtain $\sup_{n} \Vert \bm S_n \Vert < \infty$ almost surely. The proof is hence complete. 
\end{proof}

\section{Proof of Theorem \ref{thm:main-result}}\label{sec:proof-thm-mainresult}
\begin{proof}{Proof of Theorem \ref{thm:main-result}.}
	The proof is based on a sample-path-wise argument. 
	For $n \geq 1$, we first rewrite the dynamics \eqref{eq:state-dynamics} to the following recursion form
	\begin{align}
		\bm S_{n}-\bm S^*=(1-\bm \gamma_{n}) \circ (\bm S_{n-1}-\bm S^*)+\bm \gamma_{n} \circ \left(F(\bm S_{n-1})-\bm S^*+\bar{\bm{\xi}}_{n}\right),\label{eq:state-dynamics-new}
	\end{align}
	where $\bm S^*$ is defined in \eqref{def:S-star-mul}.
	Denote $S_{k,n}$ as the $k$-th entry of $\bm S_n$ ($\gamma_{k,n}$ is the $k$-th entry of $\bm \gamma_n$). We state three preliminary lemmas, the proofs of which are given in Appendix \ref{proof-lemma-Sn-recursion}, \ref{proof-lemma-stepsize} and \ref{proof-lemma-stepsize-prod}.
	\begin{lemma}\label{lemma:Sn_recursion}
		For $k = 1,\dots,2(I+J)$, let $\prod_{s=n+1}^{n} (1-\gamma_{k,s})=1$ by convention, then for any $0 \leq m \leq n$, $S_{k,n}- S^*_k$ has the following recursive form almost surely:
		\begin{align}
			S_{k,n}-S^*_k&= (S_{k,m}-S_k^*) \cdot \prod_{\tau=m+1}^{n} (1-\gamma_{k,\tau})+\sum_{\tau=m+1}^{n} \left[\prod_{s=\tau+1}^{n} (1-\gamma_{k,s})\right]\gamma_{k,\tau}\left(F_k(\bm S_{\tau-1})-S_k^*+\bar{\xi}_{k,\tau}\right).\label{recursion:Sn}
		\end{align}	
	\end{lemma}
	
	\begin{lemma}\label{lemma:stepsize}
		For $k = 1,\dots,2(I+J)$, let $\prod_{s=n+1}^{n} (1-\gamma_{k,s})=1$ by convention, then for any $1 \leq m \leq n$, we have almost surely,
		\begin{align}
			\prod_{\tau=m}^{n} (1-\gamma_{k,\tau})+ \sum_{\tau=m}^{n} \left[\prod_{s=\tau+1}^{n} (1-\gamma_{k,s})\right] \gamma_{k,\tau}=1. 
		\end{align}
	\end{lemma}
	
	\begin{lemma}\label{lemma:stepsize_prod}
		For $k = 1,\dots,2(I+J)$, we have $\prod_{\tau=1}^{\infty} (1-\gamma_{k,\tau})=0$ almost surely.
	\end{lemma}
	
	Now we present the proof of Theorem \ref{thm:main-result}, which builds on the proof of Theorem 3 in \cite{tsitsiklis1994asynchronous}.  We fix any sample path $\omega$ (that does not lie in the null sets in the three lemmas above) throughout the proof. For notational simplicity, we omit the specification of the path $\omega$ below.
	
	\textit{Step 1: Confining the Iterates to a Well-Behaved Region.} 
	Consider the recursion of $\bm S_n (n \geq 1)$ starting from period $0$, from Lemma \ref{lemma:Sn_recursion}, for any entry $k$, we have 
	\begin{align}
		S_{k,n}-S^*_k
		= (S_{k,0}-S_k^*) \cdot \prod_{\tau=1}^{n} (1-\gamma_{k,\tau})
		+\sum_{\tau=1}^{n} \left[\prod_{s=\tau+1}^{n} (1-\gamma_{k,s}) \right]\gamma_{k,\tau} \left(F_k(\bm S_{\tau-1})-S_k^*+\bar{\xi}_{k,\tau}\right).\label{recursion:Sn0}
	\end{align}
	For $n \geq 1$, let 
	\begin{align}
		&C_{k,n} \coloneqq (S_{k,0}-S_k^*) \cdot \prod_{\tau=1}^{n} (1-\gamma_{k,\tau}),\\
		&D_{k,n} \coloneqq \sum_{\tau=1}^{n} \left[ \prod_{s=\tau+1}^{n} (1-\gamma_{k,s}) \right]\gamma_{k,\tau}(F_k(\bm S_{\tau-1})-S_k^*),\\
		&E_{k,n} \coloneqq \sum_{\tau=1}^{n} \left[ \prod_{s=\tau+1}^{n} (1-\gamma_{k,s}) \right] \gamma_{k,\tau}\bar{\xi}_{k,\tau},
	\end{align}
	Then \eqref{recursion:Sn0} implies that 
	\begin{align}\label{eq:decomp}
		S_{k,n}-S_k^*= C_{k,n}+D_{k,n}+E_{k,n}, \quad \forall k. 
	\end{align}
	For the first term $C_{k,n}$, we can apply Lemma \ref{lemma:stepsize_prod} and obtain for any $k$,
	\begin{align}
		\lim_{n \to \infty} C_{k,n} =0.\label{limit:Ckn}
	\end{align}
	
	Next, let us consider the third term $E_{k,n}$. Recall the definition of $\bar{\bm{\xi}}_{n}$ in \eqref{eq:noise-mul}, we know $\bar{\xi}_{k,n}=0$ for $n \geq 1$, $k=I+J+1,\dots,2(I+J)$, which implies $E_{k,n}=0$ for these entries.
	Moreover, for any $1 \leq m \leq n-1$,  $E_{k,n}$  has the following recursion 
	\begin{align}
		E_{k,n}= \prod\limits_{\tau=m+1}^{n} (1-\gamma_{k,\tau})\cdot E_{k,m}
		+\sum\limits_{\tau=m+1}^{n}  \left[\prod\limits_{s=\tau+1}^{n}(1-\gamma_{k,s})\right] \gamma_{k,\tau} \bar{\xi}_{k,\tau}.
	\end{align}
	From the proof of Lemma 2 in \cite{tsitsiklis1994asynchronous}, we immediately have for $k=1\dots,I+J$
	\begin{align}
		\lim_{n \to \infty} E_{k,n} =0.\label{limit:Ekn}
	\end{align}
	
	Finally, we discuss the remaining second term $D_{k,n}$. For entries $k=I+J+1,\dots,2(I+J)$, by the definition of $F(\bm S)$ in \eqref{eq:F-mul}, we know $F_k(\bm S_n)=0$ for $n \geq 1$. Moreover, we know $S_k^*=0$ from \eqref{def:S-star-mul}, so $D_{k,n}=0$ for $k=I+J+1,\dots,2(I+J), n \geq 1$. Therefore, we only need to consider $D_{k,n}$ for $k=1,\dots,I+J$.
	
	Note that for $i=1,\dots,I$, we have
	\begin{align}
		D_{i,n}
		&= \sum_{\tau=1}^{n} \left[ \prod_{s=\tau+1}^{n} (1-\gamma_{i,s}) \right]\gamma_{i,\tau}\left(F_i(\bm S_{\tau-1})-S_i^*\right)\\
		&= \sum_{\tau=1}^{n} \left[ \prod_{s=\tau+1}^{n} (1-\gamma_{i,s}) \right]\gamma_{i,\tau} \left(\sum_{j=1}^J A_{i,j}\psi_{j,\tau}-A_{i,1} \right)\\
		&=\sum_{\tau=1}^{n} \left[\prod_{s=\tau+1}^{n} (1-\gamma_{i,s}) \right]\gamma_{i,\tau} \left[\sum_{j=2}^J (A_{i,j}-A_{i,1}) \psi_{j,\tau} \right].
	\end{align}
	It follows that 
	\begin{align}
		\vert D_{i,n} \vert  &\leq \max_{j=2,\dots,J} \vert A_{i,j}-A_{i,1} \vert \cdot \sum_{\tau=1}^{n} \prod_{s=\tau+1}^{n} (1-\gamma_{i,s})\gamma_{i,\tau}\\
		&= \max_{j=2,\dots,J} \vert A_{i,j}-A_{i,1} \vert \cdot \left [1-\prod_{\tau=1}^{n} (1-\gamma_{i,\tau})\right]\\
		&\leq \max_{j=2,\dots,J} \vert A_{i,j}-A_{i,1} \vert,\label{bound:Din-mul}
	\end{align}
	where the equation holds due to Lemma \ref{lemma:stepsize}.
	
	Similarly, for $j=1,\dots, J$, we have
	\begin{align}
		&\vert D_{I+j,n} \vert \leq \max_{i=2,\dots, I} \vert B_{i,j}-B_{1,j}\vert.\label{bound:Djn-mul}
	\end{align}
	
	From Assumption \ref{asp:payoff-matrix-mul}, we can obtain there exists $\epsilon_1 \in (0,1)$ and $\epsilon_2 \in (0,1)$ such that
	\begin{align}
		&\max_{j \ne 1} \vert A_{i,1}-A_{i,j} \vert + \max_{j \ne 1} \vert A_{\ell,1}-A_{\ell,j} \vert  \leq (1-\epsilon_1) \vert A_{i,1}-A_{\ell,1} \vert, \quad \forall i \ne \ell.\\
		&\max_{i \ne 1} \vert B_{1,j}-B_{i,j} \vert + \max_{i \ne 1} \vert B_{1,\ell}-B_{i,\ell} \vert \leq (1-\epsilon_2) \vert B_{1,j}-B_{1,\ell} \vert, \quad \forall j \ne \ell.\label{inequ:assum2}
	\end{align}
	
	Recall $\lim_{n \to \infty} C_{k,n} =0$ for all $k$ in \eqref{limit:Ckn}.
	Given $0<\epsilon_3<\frac{\min\{\epsilon_1, \epsilon_2\}}{2}$, we can obtain that there exists $n_0$ such that for $n>n_0$, there holds
	\begin{align}
		&\vert C_{i,n}\vert \leq \frac{\epsilon_3}{4} \cdot \min_{\ell \ne i} \vert A_{i,1}-A_{\ell,1} \vert, \quad \textrm{for $i=1,\dots,I$.}\\
		&\vert C_{I+j,n}\vert \leq \frac{\epsilon_3}{4} \cdot \min_{\ell \ne j} \vert B_{1,j}-B_{1,\ell} \vert, \quad \textrm{for $j=1,\dots,J$.}\\
		&\vert C_{I+J+i,n}\vert \leq \frac{\epsilon_3}{4} \cdot \min_{\ell \ne i} \vert A_{i,1}-A_{\ell,1} \vert, \quad \textrm{for $i=1,\dots,I$.}\\
		&\vert C_{2I+J+j,n}\vert \leq \frac{\epsilon_3}{4} \cdot \min_{\ell \ne j} \vert B_{1,j}-B_{1,\ell} \vert, \quad \textrm{for $j=1,\dots,J$.}\label{n0_condition_mul}
	\end{align}
	
	From \eqref{limit:Ekn}, we have $\lim_{n \to \infty} E_{k,n} =0$ for $k=1,\dots,I+J$.
	So there exists $\tau_0$ such that for $n>\tau_{0}$, there holds
	\begin{align}
		&\vert E_{i,n}\vert \leq \frac{\epsilon_3}{4} \cdot \min_{\ell \ne i} \vert A_{i,1}-A_{\ell,1} \vert, \: \textrm{for $i=1,\dots,I$.}\\
		&\vert E_{I+j,n}\vert \leq \frac{\epsilon_3}{4} \cdot \min_{\ell \ne j} \vert B_{1,j}-B_{1,\ell} \vert, \: \textrm{for $j=1,\dots,J$.} \label{tau0_condition_mul}
	\end{align}
	
	Therefore, for $i=1,\dots,I$, $n > \max\{n_0, \tau_0\}$, 
	we can infer from \eqref{eq:decomp}, \eqref{inequ:assum2},  \eqref{n0_condition_mul} and \eqref{tau0_condition_mul} that
	\begin{align}
		\vert x_{i,n}-A_{i,1} \vert \leq \vert C_{i,n}\vert + \vert D_{i,n}\vert+\vert E_{i,n}\vert \leq \frac{\epsilon_3}{2} \min_{\ell \ne i} \vert A_{i,1}-A_{\ell,1} \vert + \max_{j \ne 1} \vert A_{i,j}-A_{i,1} \vert.
	\end{align}
	
	Similarly, for $i=1,\dots,J$, $n > \max\{n_0, \tau_0\}$, we also have
	\begin{align}
		&\vert y_{j,n}-B_{1,j} \vert \leq \frac{\epsilon_3}{2} \min_{\ell \ne j} \vert B_{1,j}-B_{1,\ell} \vert+ \max_{i \ne 1} \vert B_{i,j}-B_{1,j}\vert.
	\end{align}	
	
	Then for $n > \max\{n_0, \tau_0\}$, we want to analyze the difference between two posterior mean $x_{i,n}-x_{k,n}$ for $i \ne k$. The argument is structured around two separate cases. If $A_{i,1} > A_{k,1}$, we have
	\begin{align}
		x_{i,n}-x_{k,n}
		& \geq A_{i,1}- \frac{\epsilon_3}{2} \min_{\ell \ne i} \vert A_{i,1}-A_{\ell,1} \vert -\max_{j \ne 1} \vert A_{i,j}-A_{i,1} \vert \\
		&\quad -\left[A_{k,1}+\frac{\epsilon_3}{2} \min_{\ell \ne k} \vert A_{k,1}-A_{\ell,1} \vert+\max_{j \ne 1} \vert A_{k,j}-A_{k,1} \vert  \right].
	\end{align}
	From \eqref{inequ:assum2}, we can further lower bound the above formula by
	\begin{align}
		&x_{i,n}-x_{k,n}\\
		&\geq (1-\epsilon_3) (A_{i,1}-A_{k,1}) -(1-\epsilon_1) (A_{i,1}-A_{k,1})\\
		&=(\epsilon_1-\epsilon_3) (A_{i,1}-A_{k,1})\\
		&>\frac{\epsilon_1}{2}(A_{i,1}-A_{k,1}).\label{inequ:x-diff-1}
	\end{align}
	If $A_{k,1}>A_{i,1}$, we have 
	\begin{align}
		&x_{k,n}-x_{i,n}\\
		& \geq A_{k,1}-\frac{\epsilon_3}{2} \min_{\ell \ne k} \vert A_{k,1}-A_{\ell,1} \vert-\max_{j \ne 1} \vert A_{k,j}-A_{k,1} \vert\\
		&\quad -\left[A_{i,1} + \frac{\epsilon_3}{2} \min_{\ell \ne i} \vert A_{i,1}-A_{\ell,1} \vert + \max_{j \ne 1} \vert A_{i,j}-A_{i,1} \vert \right]\\
		&\geq (1-\epsilon_3)(A_{k,1}-A_{i,1}) -(1-\epsilon_1) (A_{k,1}-A_{i,1})\\
		&> \frac{\epsilon_1}{2} (A_{k,1}-A_{i,1}).\label{inequ:x-diff-2}
	\end{align}
	Combining the above two cases, we have 
	\begin{align}
		\vert x_{i,n}-x_{k,n} \vert >\frac{\epsilon_1}{2}\vert A_{i,1}-A_{k,1}\vert >0.
	\end{align}
	Similarly, for $n > \max\{n_0, \tau_0\}$, there holds
	\begin{align}
		\vert y_{j,n}-y_{k,n} \vert > \frac{\epsilon_2}{2}\vert B_{1,j}-B_{1,k}\vert>0.
	\end{align}	
	
	What we have shown is that $\bm S_n$ will avoid the region where $F$ is not Lipschitz continuous for a sufficiently large $n$. 
	
	\textit{Step 2: Establishing a Local Contraction Property.} 
	We next show that there exists $\beta \in [0,1)$ such that for $n$ large enough,
	\begin{align}
		\Vert F(\bm S_n) -F(\bm S^*)\Vert_\infty \leq \beta \Vert \bm S_n -\bm S^* \Vert_\infty.
	\end{align}
	
	To prove the results, we apply the mean value theorem. We have for $i=1,\dots,I+J$,
	\begin{align}\label{mean-value-thm}
		F_i (\bm S_n) -F_i (\bm S^*) = \nabla F_i (\tilde{\bm S}_n) \cdot ( \bm S_n -\bm S^*),
	\end{align}
	where $\tilde{\bm S}_n=(\tilde{x}_{1,n}, \dots, \tilde{x}_{I,n}, \tilde{y}_{1,n}, \dots, \tilde{y}_{J,n}, \tilde{w}_{1,n}, \dots, \tilde{w}_{I,n}, \tilde{v}_{1,n}, \dots, \tilde{v}_{J,n})$ is a point on the line segment between $\bm S_{n}$ and $\bm S^*$. 
	Hence it suffices to bound the gradient $\nabla F_i (\tilde{\bm S}_n)$.
	Write the Jacobian matrix  
	\begin{align}
		L=\begin{pmatrix}
			L_{1,1} &L_{1,2} &\cdots &L_{1,2(I+J)}\\
			L_{2,1} &L_{2,2} &\cdots &L_{2,2(I+J)}\\
			\vdots &\vdots &\ddots &\vdots\\
			L_{I+J,1} &L_{I+J,2} &\cdots &L_{I+J,2(I+J)}
		\end{pmatrix} \in \mathbb{R}^{(I+J) \times 2(I+J)},
	\end{align}
	where $L_{\ell,k}=\frac{\partial F_\ell(\bm S)}{\partial S_k}|_{\bm S=\tilde{\bm S}_n}, \ell=1,\dots,I+J, k=1,\dots,2(I+J)$.
	
	By the definition of $F$ in \eqref{eq:F-mul}, it is easy to see that
	\begin{align}
		&L_{i,1}=\dots=L_{i,I}=0,\quad\forall i=1,\dots,I.\\
		&L_{i,I+J+1}=\dots=L_{i,2I+J}=0,\quad\forall i=1,\dots,I.\\
		&L_{I+j,I+1}=\dots=L_{I+j,I+J}= 0,\quad\forall j=1,\dots,J.\\
		&L_{I+j,2I+J+1}=\dots=L_{I+j,2(I+J)}= 0,\quad\forall j=1,\dots,J.
	\end{align}
	
	To analyze other terms of $L_{\ell,k}$, we need to compute the derivatives of the functions $F_\ell$, we begin by calculating the gradients of $\varphi_{i}$ and $\psi_{j}$. For instance, from \eqref{notation:phi-mul}, we have the derivative of $\varphi_{i}$ with respect to $x_i$:
	\begin{align}
		\frac{\partial \varphi_{i}}{\partial x_{i}}=\frac{\partial \Phi_{I-1}\left(\left(c_{k}^{(i)}\right)_{k \ne i}; \rho_{i}\right)}{\partial x_{i}}
		=\sum_{k \ne i}\frac{\partial \Phi_{I-1}\left(\left(c_{k}^{(i)}\right)_{k \ne i}; \rho_{i}\right)}{\partial c_{k}^{(i)}} \cdot \frac{\partial c_{k}^{(i)}}{\partial x_{i}}.\label{eq:phi-mul-der}
	\end{align}
	For the first term, we calculate the gradient of the distribution function of the multivariate standard normal distribution (Section 6.6.4, \cite{prekopa2013stochastic}):
	\begin{align}
		\frac{\partial \Phi_{I-1}\left(\left(c_{k}^{(i)}\right)_{k \ne i}; \rho_{i}\right)}{\partial c_{k}^{(i)}}
		=\Phi_{I-2} \left(\left(c_{\ell}^{(i)}\right)_{\ell \ne i,k} \big |c_{k}^{(i)} \right) \cdot \phi \left(c_{k}^{(i)}\right),
	\end{align}
	where $\phi(z)$ is the probability density function of the standard normal distribution, and $\Phi_{I-2} \left(\left(c_{\ell}^{(i)}\right)_{\ell \ne i,k} \big |c_{k}^{(i)} \right)$ is the conditional distribution function of the normal random variables $\left(Z_{\ell}^{(i)}\right)_{\ell \ne k,i}$ given $Z_{k}^{(i)}=c_{k}^{(i)}$.\\
	For the second term, recall the definition of $c_{k}^{(i)}$ as $\frac{x_{i}-x_{k}}{\sqrt{w_{i}+w_{k}}}$ for $k \ne i$ , we have 
	\begin{align}
		\frac{\partial c_{k}^{(i)}}{\partial x_{i}}=\frac{1}{\sqrt{w_{i}+w_{k}}}.
	\end{align}
	Plugging them into \eqref{eq:phi-mul-der}, we have 
	\begin{align}
		\frac{\partial \varphi_{i}}{\partial x_{i}}=\sum_{k \ne i}\Phi_{I-2} \left(\left(c_{\ell}^{(i)}\right)_{\ell \ne i,k} \big |c_{k}^{(i)} \right) \cdot \frac{\phi \left(c_{k}^{(i)}\right)}{\sqrt{w_{i}+w_{k}}}.\label{eq:phi-mul-der-p1}
	\end{align}
	
	Similarly, we can obtain other derivatives of $\varphi_i$:
	\begin{align}
		&\frac{\partial \varphi_{i}}{\partial x_{k}}=-\Phi_{I-2} \left(\left(c_{\ell}^{(i)}\right)_{\ell \ne i,k} \big |c_{k}^{(i)} \right) \cdot \frac{\phi \left(c_{k}^{(i)}\right)}{\sqrt{w_{i}+w_{k}}},\quad \forall k \ne i,\\
		&\frac{\partial \varphi_{i}}{\partial w_{i}}=\sum_{k \ne i}\Phi_{I-2} \left(\left(c_{\ell}^{(i)}\right)_{\ell \ne i,k} \big |c_{k}^{(i)} \right) \cdot \phi \left(c_{k}^{(i)}\right) \cdot \frac{x_{k}-x_{i}}{2(w_{i}+w_{k})^{3/2}},\\
		&\frac{\partial \varphi_{i}}{\partial w_{k}}
		=\Phi_{I-2} \left(\left(c_{\ell}^{(i)}\right)_{\ell \ne i,k} \big |c_{k}^{(i)} \right) \cdot \phi \left(c_{k}^{(i)}\right) \cdot \frac{x_{k}-x_{i}}{2(w_{i}+w_{k})^{3/2}}, \quad \forall k \ne i.\label{eq:phi-mul-der-p2}
	\end{align}
	
	For $j=1,\dots,J$, we can also obtain the following gradients of $\psi_{j}$.
	\begin{align}
		&\frac{\partial \psi_{j}}{\partial y_{j}}=\sum_{k \ne j}\Phi_{J-2} \left(\left(d_{\ell}^{(j)}\right)_{\ell \ne j,k} \big |d_{k}^{(j)} \right) \cdot \frac{\phi \left(d_{k}^{(j)}\right)}{\sqrt{v_{j}+v_{k}}},\\
		&\frac{\partial \psi_{j}}{\partial y_{k}}=-\Phi_{J-2} \left(\left(d_{\ell}^{(j)}\right)_{\ell \ne j,k} \big |d_{k}^{(j)} \right) \cdot \frac{\phi \left(d_{k}^{(j)}\right)}{\sqrt{v_{j}+v_{k}}},\quad \forall k \ne j,\\
		&\frac{\partial \psi_{j}}{\partial v_{j}}=\sum_{k \ne j}\Phi_{J-2} \left(\left(d_{\ell}^{(j)}\right)_{\ell \ne j,k} \big |d_{k}^{(j)} \right) \cdot \phi \left(d_{k}^{(j)}\right) \cdot \frac{y_{k}-y_{j}}{2(v_{j}+v_{k})^{3/2}},\\
		&\frac{\partial \psi_{j}}{\partial v_{k}}=\Phi_{J-2} \left(\left(d_{\ell}^{(j)}\right)_{\ell \ne j,k} \big |d_{k}^{(j)} \right) \cdot \phi \left(d_{k}^{(j)}\right) \cdot \frac{y_{k}-y_{j}}{2(v_{j}+v_{k})^{3/2}}, \quad \forall k \ne j,\label{eq:psi-mul-der}
	\end{align}
	where $d_{k}^{(j)}=\frac{y_{j}-y_{k}}{\sqrt{v_{j}+v_{k}}}$ for $k \ne j$.
	
	So, the formulas of $L_{\ell,k}$ can be derived. For example,
	\begin{align}
		L_{i,I+j} &=\frac{\partial F_{i}}{\partial y_j}\Big |_{\bm S = \tilde{\bm S}_n}=\frac{\partial \sum_{k=1}^J A_{i,k} \psi_{k}}{\partial y_{j}}\Big |_{\bm S = \tilde{\bm S}_n}
		=A_{i,j}\frac{\partial \psi_{j}}{\partial y_{j}} \Big |_{\bm S = \tilde{\bm S}_n}+\sum_{k \ne j} A_{i,k}\frac{\partial \psi_{k}}{\partial y_{j}} \Big |_{\bm S = \tilde{\bm S}_n}.
	\end{align}
	Substituting the gradients from \eqref{eq:psi-mul-der} into the formula, we get
	\begin{align}
		L_{i,I+j}
		&=A_{i,j} \left[\sum_{k \ne j}\Phi_{J-2} \left(\left(\tilde{d}_{\ell,n}^{(j)}\right)_{\ell \ne j,k} \big |\tilde{d}_{k,n}^{(j)} \right) \cdot \frac{\phi \left(\tilde{d}_{k,n}^{(j)}\right)}{\sqrt{\tilde{v}_{j,n}+\tilde{v}_{k,n}}}\right]\\
		&\quad-\sum_{k \ne j} A_{i,k} \Phi_{J-2} \left(\left(\tilde{d}_{\ell,n}^{(k)}\right)_{\ell \ne k,j} \big |\tilde{d}_{j,n}^{(k)} \right) \cdot \frac{\phi \left(\tilde{d}_{j,n}^{(k)}\right)}{\sqrt{\tilde{v}_{k,n}+\tilde{v}_{j,n}}}\\
		&=\sum_{ k \ne j} \frac{\phi \left(\tilde{d}_{k,n}^{(j)}\right)}{\sqrt{\tilde{v}_{j,n}+\tilde{v}_{k,n}}} \cdot \left[A_{i,j}\Phi_{J-2} \left(\left(\tilde{d}_{\ell,n}^{(j)}\right)_{\ell \ne j,k} \big |\tilde{d}_{k,n}^{(j)} \right)-A_{i,k}\Phi_{J-2} \left(\left(\tilde{d}_{\ell,n}^{(k)}\right)_{\ell \ne k,j} \big |\tilde{d}_{j,n}^{(k)} \right) \right].
	\end{align}
	Likewise, we can derive the following formulas.
	\begin{align}
		&L_{i,2I+J+j} 
		=\sum_{ k \ne j} \frac{\phi \left(\tilde{d}_{k,n}^{(j)}\right) \cdot (\tilde{y}_{k,n}-\tilde{y}_{j,n})}{\sqrt{\tilde{v}_{j,n}+\tilde{v}_{k,n}}} \cdot \left[A_{i,j}\Phi_{J-2} \left(\left(\tilde{d}_{\ell,n}^{(j)}\right)_{\ell \ne j,k} \big |\tilde{d}_{k,n}^{(j)} \right)-A_{i,k}\Phi_{J-2} \left(\left(\tilde{d}_{\ell,n}^{(k)}\right)_{\ell \ne k,j} \big |\tilde{d}_{j,n}^{(k)} \right) \right],\\
		&L_{I+j,i}
		=\sum_{k \ne i} \frac{\phi \left(\tilde{c}_{k,n}^{(i)}\right)}{\sqrt{\tilde{w}_{i,n}+\tilde{w}_{k,n}}} \left[B_{i,j} \Phi_{I-2} \left(\left(\tilde{c}_{\ell,n}^{(i)}\right)_{\ell \ne i,k} \big |\tilde{c}_{k,n}^{(i)} \right)- B_{k,j}\Phi_{I-2} \left(\left(\tilde{c}_{\ell,n}^{(k)}\right)_{\ell \ne k,i} \big |\tilde{c}_{i,n}^{(k)} \right) \right],\\
		&L_{I+j,I+J+i}
		=\sum_{k \ne i} \frac{\phi \left(\tilde{c}_{k,n}^{(i)}\right) \cdot (\tilde{x}_{k,n}-\tilde{x}_{i,n})}{\sqrt{\tilde{w}_{i,n}+\tilde{w}_{k,n}}} \cdot \left[B_{i,j}\Phi_{I-2} \left(\left(\tilde{c}_{\ell,n}^{(i)}\right)_{\ell \ne i,k} \big |\tilde{c}_{k,n}^{(i)} \right)-B_{k,j}\Phi_{I-2} \left(\left(\tilde{c}_{\ell,n}^{(k)}\right)_{\ell \ne k,i} \big |\tilde{c}_{i,n}^{(k)} \right) \right].
	\end{align}
	
	To bound these $L_{\ell,k}$ terms, we recall the definition of $\bm S^*$ in \eqref{def:S-star-mul}, and note that there exists $\zeta \in (0,1)$, such that $\tilde{\bm S}_n=\zeta \bm S_{n}+(1-\zeta) \bm S^*$.
	Then for $n > \max\{n_0, \tau_0\}$, we can infer that
	\begin{align}
		&\tilde{w}_{i,n}+\tilde{w}_{k,n}=\zeta(w_{i,n}+w_{k,n})+(1-\zeta)\cdot 0\leq w_{i,n}+w_{k,n},\\
		&\tilde{v}_{j,n}+\tilde{v}_{k,n}=\zeta(v_{j,n}+v_{k,n})+(1-\zeta)\cdot 0\leq v_{j,n}+v_{k,n}.
	\end{align}
	
	To analyze $\vert \tilde{x}_{i,n}-\tilde{x}_{k,n} \vert$, we divide the discussion into the following two cases.
	If $A_{i,1} >A_{k,1}$, from \eqref{inequ:x-diff-1}, we know
	\begin{align}
		\tilde{x}_{i,n}-\tilde{x}_{k,n}
		=\zeta(x_{i,n}-x_{k,n})+(1-\zeta) (A_{i,1}-A_{k,1})
		> \frac{\epsilon_1}{2} (A_{i,1}-A_{k,1}).
	\end{align}
	If $A_{k,1} >A_{i,1}$, from \eqref{inequ:x-diff-2}, we have
	\begin{align}
		\tilde{x}_{k,n}-\tilde{x}_{i,n}
		=\zeta(x_{k,n}-x_{i,n})+(1-\zeta) (A_{k,1}-A_{i,1})
		> \frac{\epsilon_1}{2} (A_{k,1}-A_{i,1}).
	\end{align}
	We combine the above two cases to get
	\begin{align}
		\vert \tilde{x}_{i,n}-\tilde{x}_{k,n} \vert \geq \frac{\epsilon_1}{2} \vert A_{i,1}-A_{k,1}\vert.
	\end{align}
	
	Similarly, we have 
	\begin{align}
		&\vert \tilde{y}_{j,n}-\tilde{y}_{k,n} \vert  > \frac{\epsilon_2}{2} \vert B_{1,j}-B_{1,k} \vert.
	\end{align}
	
	Therefore, we can bound $L_{\ell,k}$ terms because the range of the distribution function is $[0,1]$.
	\begin{align}
		\left \vert L_{i,I+j} \right\vert
		&\leq \sum_{ k \ne j} \frac{1}{\sqrt{\tilde{v}_{j,n}+\tilde{v}_{k,n}}} \cdot \frac{1}{\sqrt{2 \pi}} \exp \left[-\frac{1}{2}\left(\tilde{d}_{k,n}^{(j)}\right)^2 \right] \cdot A_{i,j}\\
		&\leq \frac{A_{i,j}}{\sqrt{2 \pi}} \sum_{ k \ne j} e^{-\frac{1}{2} \left(\tilde{d}_{k,n}^{(j)}\right)^2}\cdot \left\vert \tilde{d}_{k,n}^{(j)} \right\vert \cdot \frac{1}{\vert \tilde{y}_{j,n}-\tilde{y}_{k,n} \vert}\\
		&\leq \frac{A_{i,j}}{\sqrt{2 \pi}} \sum_{ k \ne j}e^{-\frac{1}{2} \left(\tilde{d}_{k,n}^{(j)}\right)^2}\cdot \left\vert \tilde{d}_{k,n}^{(j)} \right \vert \cdot\frac{2}{\epsilon_2 \vert B_{1,j}-B_{1,k}\vert},
	\end{align}
	and
	\begin{align}
		\left \vert L_{i,2I+J+j} \right \vert
		&\leq \sum_{ k \ne j} \frac{\tilde{y}_{k,n}-\tilde{y}_{j,n}}{\sqrt{\tilde{v}_{j,n}+\tilde{v}_{k,n}}} \cdot \frac{1}{\sqrt{2 \pi}} \exp \left[-\frac{1}{2}\left(\tilde{d}_{k,n}^{(j)}\right)^2 \right] \cdot A_{ij}\\
		&\leq \frac{A_{i,j}}{\sqrt{2 \pi}} \sum_{ k \ne j} e^{-\frac{1}{2} \left(\tilde{d}_{k,n}^{(j)}\right)^2}\cdot \left \vert \tilde{d}_{k,n}^{(j)} \right \vert^3 \cdot \frac{1}{\vert \tilde{y}_{j,n}-\tilde{y}_{k,n} \vert ^2}\\
		&\leq \frac{A_{i,j}}{\sqrt{2 \pi}} \sum_{ k \ne j} e^{-\frac{1}{2} \left(\tilde{d}_{k,n}^{(j)}\right)^2}\cdot \left \vert \tilde{d}_{k,n}^{(j)} \right \vert^3 \cdot \frac{4}{\epsilon_2^2 \vert B_{1,j}-B_{1,k}\vert^2}.
	\end{align}
	
	Similarly,
	\begin{align}
		&\left \vert L_{I+j,i} \right \vert \leq \frac{B_{i,j}}{\sqrt{2 \pi}} \sum_{k \ne i}e^{-\frac{1}{2} \left(\tilde{c}_{k,n}^{(i)}\right)^2}\cdot \left \vert \tilde{c}_{k,n}^{(i)} \right \vert \cdot\frac{2}{\epsilon_1 \vert A_{i,1}-A_{k,1}\vert},\\
		&\left \vert L_{i,2I+J+j} \right \vert \leq \frac{B_{i,j}}{\sqrt{2 \pi}} \sum_{k \ne i} e^{-\frac{1}{2} \left(\tilde{c}_{k,n}^{(i)}\right)^2}\cdot \left \vert \tilde{c}_{k,n}^{(i)} \right \vert^3 \cdot \frac{4}{\epsilon_1^2 \vert A_{i,1}-A_{k,1}\vert^2}.
	\end{align}
	
	Let $h_1(z)=e^{-\frac{1}{2}z^2}\cdot z$, which is a decreasing function when $z \in [1,\infty)$ and $z \in (-\infty,-1]$. Let $h_2(z)=e^{-\frac{1}{2}z^2}\cdot z^3$, which is also a decreasing function when $z \in [\sqrt{3},\infty)$ and $z \in (-\infty,-\sqrt{3}]$. 
	And we have
	\begin{align}
		&\left \vert \tilde{c}_{k,n}^{(i)} \right \vert = \frac{\vert \tilde{x}_{i,n}-\tilde{x}_{k,n}\vert }{\sqrt{\tilde{w}_{i,n}+\tilde{w}_{k,n}}}
		\geq \frac{\epsilon_1 \vert A_{i,1}-A_{k,1} \vert}{2 \sqrt{w_{i,n}+w_{k,n}}}
		=\frac{\epsilon_1\vert A_{i,1}-A_{k,1}\vert}{2\sqrt{\frac{1}{N_{i,n}+1}+\frac{1}{N_{k,n}+1}}},\\
		&\left \vert \tilde{d}_{k,n}^{(j)} \right \vert = \frac{\vert \tilde{y}_{j,n}-\tilde{y}_{k,n}\vert }{\sqrt{\tilde{v}_{j,n}+\tilde{v}_{k,n}}}\geq \frac{\epsilon_2 \vert B_{1,j}-B_{1,k} \vert}{2 \sqrt{v_{j,n}+v_{k,n}}}=\frac{\epsilon_2\vert B_{1,j}-B_{1,k}\vert}{2\sqrt{\frac{1}{M_{j,n}+1}+\frac{1}{M_{k,n}+1}}}.
	\end{align}
	
	From Lemma \ref{lemma:infinite}, we have $\lim_{n \to \infty} N_{i,n}=\infty$ ($i=1,\dots,I$) and $\lim_{n \to \infty} M_{j,n}=\infty$ ($j=1,\dots,J$) almost surely. We then let $\bar{L}_\ell \coloneqq \max_{k=1,\dots,2(I+J)} \vert L_{\ell,k} \vert$, $\ell=1,\dots,I+J$.  We can infer that there exists $\eta_0>\max\{n_0, \tau_0\}$ such that for $n > \eta_0$, 
	\begin{align}
		\bar{L}_\ell \leq \frac{1}{4(I+J)}<1.
	\end{align}
	
	\textit{Step 3. Obtain the convergence of $\bm S_n$ to $\bm S^*$}. Finally, by the mean value theorem, for $n > \eta_0$, we have
	\begin{align}
		&\vert F_\ell(\bm S_n) -F_\ell(\bm S^*)\vert
		\leq \bar{L}_\ell \sum_{k=1}^{2(I+J)}  \vert S_{k,n}-S_k^* \vert
		\leq 2(I+J) \bar{L}_\ell \Vert \bm S_n -\bm S^* \Vert_\infty
		\leq \frac{1}{2}\Vert \bm S_n -\bm S^* \Vert_\infty.
	\end{align}
	Therefore, we obtain for $n>\eta_0$,
	\begin{align}
		\Vert F(\bm S_n) -F(\bm S^*)\Vert_\infty \leq \frac{1}{2} \Vert \bm S_n -\bm S^* \Vert_\infty.
	\end{align} 
	Following the proof of Theorem 3 in \cite{tsitsiklis1994asynchronous}, we have that $\bm S_n$ converges to $\bm S^*$. 
\end{proof}

\subsection{Proof of Lemma \ref{lemma:Sn_recursion}}\label{proof-lemma-Sn-recursion}
\begin{proof}{Proof of Lemma \ref{lemma:Sn_recursion}.}
	We prove this lemma by induction. Firstly, when $n=m$, this recursion obviously holds. Suppose it holds for time $n=N$, which means that 
	\begin{small}
		\begin{align}
			S_{k,N}-S_k^* = (S_{k,m}-S_k^*) \cdot \prod_{\tau=m+1}^{N} (1-\gamma_{k,\tau}) +\sum_{\tau=m+1}^{N} \left(\prod_{s=\tau+1}^{N} (1-\gamma_{k,s})\right)\gamma_{k,\tau}\left(F_k(\bm S_{\tau-1})-S_k^*+\bar{\xi}_{k,\tau}\right).\label{forluma:SkN-Skstar}
		\end{align}
	\end{small}
	Next consider time $n=N+1$.
	From \eqref{eq:state-dynamics-new}, we first have
	\begin{align}
		S_{k,N+1}-S^*_k=(1-\gamma_{k,N+1}) (S_{k,N}-S_k^*)+\gamma_{k,N+1} \left(F_k(\bm S_{N})- S_k^*+\bar{\xi}_{k,N+1}\right).
	\end{align}
	Based on the assumption for $n=N$, we then replace term $S_{k,N}-S_k^*$ by right hand side of \eqref{forluma:SkN-Skstar}:
	\begin{small}
		\begin{align}
			&S_{k,N+1}-S^*_k\\
			&=(1-\gamma_{k,N+1}) (S_{k,N}-S_k^*)+\gamma_{k,N+1} \left(F_k(\bm S_{N})- S_k^*+\bar{\xi}_{k,N+1}\right)\\
			&=(1-\gamma_{k,N+1}) \cdot \left[(S_{k,m}-S_k^*) \cdot \prod_{\tau=m+1}^{N} (1-\gamma_{k,\tau}) +\sum_{\tau=m+1}^{N} \left(\prod_{s=\tau+1}^{N} (1-\gamma_{k,s})\right)\gamma_{k,\tau}\left(F_k(\bm S_{\tau-1})-S_k^*+\bar{\xi}_{k,\tau}\right)\right]\\
			&\quad+\gamma_{k,N+1} \left(F_k(\bm S_{N})- S_k^*+\bar{\xi}_{k,N+1}\right)\\
			&=(S_{k,m}-S_k^*) \cdot \prod_{\tau=m+1}^{N+1} (1-\gamma_{k,\tau}) + \sum_{\tau=m+1}^{N} \left[\prod_{s=\tau+1}^{N+1} (1-\gamma_{k,s}) \right]\gamma_{k,\tau}\left(F_k(\bm S_{\tau-1})-S_k^*+\bar{\xi}_{k,\tau}\right) \\
			&\quad+\gamma_{k,N+1} \left(F_k(\bm S_{N})- S_k^*+\bar{\xi}_{k,N+1}\right).\label{proof:Sn}
		\end{align}
	\end{small}
	Note that $\prod\limits_{s=N+2}^{N+1} (1-\gamma_{k,s})=1$, which implies the last term of \eqref{proof:Sn} can be rewritten
	\begin{align}
		\gamma_{k,N+1} \left(F_k(\bm S_{N})- S_k^*+\bar{\xi}_{k,N+1} \right)= \prod_{s=N+2}^{N+1} (1-\gamma_{k,s}) \cdot \gamma_{k,N+1}\left(F_k(\bm S_{N})-S_k^*+\bar{\xi}_{k,N+1}\right).
	\end{align}
	So we can further rewrite the right hand side of \eqref{proof:Sn} to
	\begin{align}
		&S_{k,N+1}-S^*_k\\
		&=(S_{k,m}-S_k^*) \cdot \prod_{\tau=m+1}^{N+1} (1-\gamma_{k,\tau}) + \sum_{\tau=m+1}^{N} \left[\prod_{s=\tau+1}^{N+1} (1-\gamma_{k,s}) \right] \gamma_{k,\tau}\left(F_k(\bm S_{\tau-1})-S_k^*+\bar{\xi}_{k,\tau}\right) \\
		&\quad+\prod_{s=N+2}^{N+1} (1-\gamma_{k,s}) \cdot \gamma_{k,N+1}\left(F_k(\bm S_{N})-S_k^*+\bar{\xi}_{k,N+1}\right)\\
		&= (S_{k,m}-S_k^*) \cdot \prod_{\tau=m+1}^{N+1} (1-\gamma_{k,\tau}) + \sum_{\tau=m+1}^{N+1} \left[\prod_{s=\tau+1}^{N+1} (1-\gamma_{k,s})\right] \gamma_{k,\tau}\left(F_k(\bm S_{\tau-1})-S_k^*+\bar{\xi}_{k,\tau}\right).
	\end{align}
	Therefore, the statement holds for time $n=N+1$, which completes the proof.
\end{proof}

\subsection{Proof of Lemma \ref{lemma:stepsize}}\label{proof-lemma-stepsize}
\begin{proof}{Proof of Lemma \ref{lemma:stepsize}.}
	We prove this lemma by induction. When $n=m$, the statement is obviously true. Suppose it is true for time $n=N$, \emph{i.e.,}
	\begin{align}
		\prod_{\tau=m}^{N} (1-\gamma_{k,\tau})+ \sum_{\tau=m}^{N} \left[\prod_{s=\tau+1}^{N} (1-\gamma_{k,s})\right] \gamma_{k,\tau}=1.
	\end{align}	
	Then consider $n=N+1$,
	\begin{align}
		&\prod_{\tau=m}^{N+1} (1-\gamma_{k,\tau})+ \sum_{\tau=m}^{N+1} \left[\prod_{s=\tau+1}^{N+1} (1-\gamma_{k,s})\right]\gamma_{k,\tau}\\
		&=(1-\gamma_{k,N+1}) \cdot \prod_{\tau=m}^{N} (1-\gamma_{k,\tau})+\sum_{\tau=m}^{N} \left[ \prod_{s=\tau+1}^{N+1} (1-\gamma_{k,s})\right] \gamma_{k,\tau} +\gamma_{k,N+1}\\
		&=(1-\gamma_{k,N+1}) \cdot \prod_{\tau=m}^{N} (1-\gamma_{k,\tau}) +(1-\gamma_{k,N+1}) \cdot \sum_{\tau=m}^{N} \left[\prod_{s=\tau+1}^{N} (1-\gamma_{k,s})\right]\gamma_{k,\tau} +\gamma_{k,N+1}\\
		&=(1-\gamma_{k,N+1}) \left[\prod_{\tau=m}^{N} (1-\gamma_{k,\tau})+ \sum_{\tau=m}^{N} \left(\prod_{s=\tau+1}^{N} (1-\gamma_{k,s})\right) \gamma_{k,\tau} \right] +\gamma_{k,N+1}\\
		&=1,
	\end{align}
	where the first equation is from $\left[\prod\limits_{s=\tau+1}^{N+1} (1-\gamma_{k,s})\right]\gamma_{k,\tau}=\gamma_{k,N+1}$ when $\tau=N+1$, and the last equality is obtained from the induction assumption for $n=N$.
	Therefore, the statement holds for time $n=N+1$, which completes the proof. 
\end{proof}

\subsection{Proof of Lemma \ref{lemma:stepsize_prod}}\label{proof-lemma-stepsize-prod}
\begin{proof}{Proof of Lemma \ref{lemma:stepsize_prod}.}
	Consider
	\begin{align}
		&\log \left[\prod_{\tau=1}^{n} (1-\gamma_{k,\tau}) \right]
		=\sum_{\tau=1}^{n} \log \left(1-\gamma_{k,\tau}\right)
		\leq -\sum_{\tau=1}^{n} \gamma_{k,\tau},
	\end{align}
	where the inequality is due to $\log x \leq x-1$ for all $x >0$.
	We know that $\sum_{n=1}^{+\infty} \gamma_{k,n}=\infty$ almost surely for all $k=1, \cdots, 2(I+J)$ from Lemma \ref{lem:step_size}. Therefore,
	\begin{align}
		\lim_{n \to \infty} \log \left[\prod_{\tau=1}^{n} (1-\gamma_{k,\tau}) \right]=-\infty, \quad a.s.
	\end{align}
	which implies $\prod\limits_{\tau=1}^{\infty} (1-\gamma_{k,\tau})=0$ almost surely. 
\end{proof}
	

\end{document}